\documentclass[aps,prb,twocolumn,superscriptaddress,showpacs,floatfix,10pt]{revtex4-1}
\usepackage{amsmath}
\usepackage{graphicx}
\usepackage{color}
\usepackage{epstopdf}
\usepackage{natbib}

\begin{document}
\title{Ultrafast magnetisation dynamics: microscopic electronic configurations and ultrafast spectroscopy}

\author{I. L. M. Locht}
\affiliation{Dept.\ of Physics and Astronomy, Uppsala University, Box 516, SE-75120 Uppsala, Sweden}
\author{I. Di Marco}
\affiliation{Dept.\ of Physics and Astronomy, Uppsala University, Box 516, SE-75120 Uppsala, Sweden}
\author{S. Garnerone}
\affiliation{Institute for Quantum Computing, University of Waterloo, Waterloo, Canada}
\author{A. Delin}
\affiliation{Dept.\ of Physics and Astronomy, Uppsala University, Box 516, SE-75120 Uppsala, Sweden}
\affiliation{Department of Materials and Nanophysics, School of Information and Communication Technology, Electrum 229, Royal Institute of Technology (KTH), SE-16440 Kista, Sweden}
\affiliation{SeRC (Swedish e-Science Research Center), KTH, SE-10044 Stockholm, Sweden}
\author{M. Battiato}
\email[]{marco.battiato@ifp.tuwien.ac.at}
\affiliation{Dept.\ of Physics and Astronomy, Uppsala University, Box 516, SE-75120 Uppsala, Sweden}
\affiliation{Institute of Solid State Physics, Vienna University of Technology,  Vienna, Austria}

\date{\today}

\begin{abstract}
We provide an approach for the identification of the electronic and magnetic configurations of ferromagnetic Fe after ultrafast decrease or increase of magnetization. The model is based on the well-grounded assumption that, after an ultrafast variation of magnetization, the system achieves a partial thermal equilibrium. With statistical arguments we show that the magnetic configurations are qualitatively different in the case of reduced or increased magnetization. The predicted magnetic configurations are then used to compute the dielectric response at the 3p (M) absorption edge, which are directly related to the changes observed in the experimental T-MOKE data. The good qualitative agreement between theory and experiment offers a substantial support to the validity of the model, and to the very existence of an ultrafast increase of magnetisation. 
\end{abstract}


\maketitle

\section{Introduction}\label{sec:Introduction}

\begin{figure}[t]
 \includegraphics[width=0.49\textwidth]{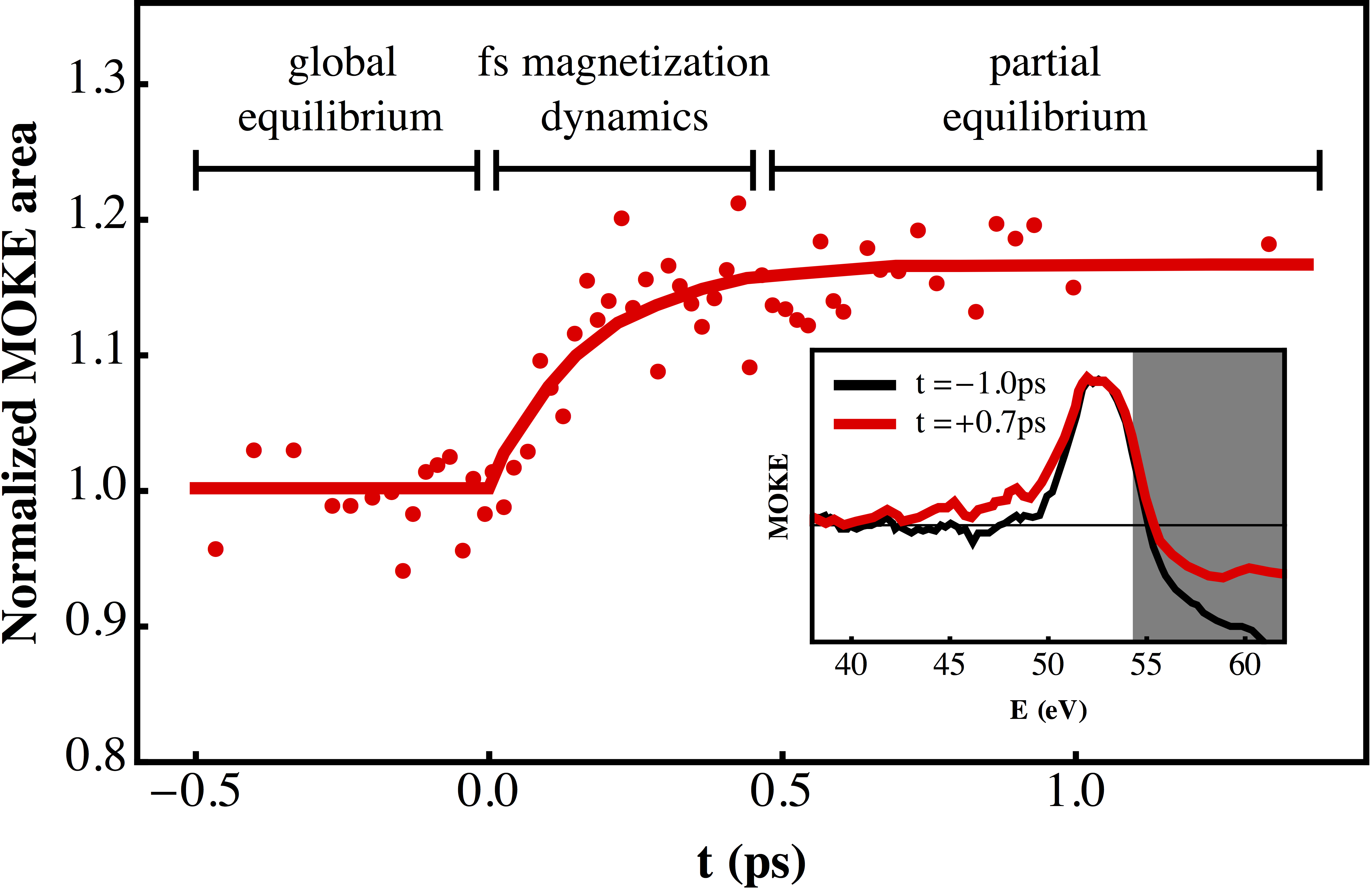}
 \caption{
Time dependent magnetic asymmetry. Time-evolution of the area spanned by the 3p asymmetry Fe peak at 53 eV as reported in Ref.~\onlinecite{Turgut13} for increased Fe magnetization. The relevant timescales in the ultrafast magnetization dynamics in this experiment are also shown (a more detailed discussion is done in Sec.~\ref{sec:equilibration}). In the inset the T-MOKE spectra at the M absorption edge are shown before (thin black line) and after (thick red line) the laser excitation, corresponding roughly to -1.0 ps and +0.7 ps\cite{Turgut13}.}\label{fig:expdata}
\end{figure}

The search for the next generation magnetic recording media is focusing on the ultrafast magnetization dynamics.\cite{Beaurepaire96,kirilyuk10,Stanciu07,Malinowski08,Vahaplar09} Despite experimental progress, \cite{stamm07,Melnikov08,mueller09,Bigot09,Radu11,Laovorakiat12,Mathias12,Ostler12,Carley12,kampfrathnatnano2013} the microscopic understanding of the ultrafast magnetization dynamics remains an open question. In the last few years, several theories were proposed as possible explanations, \cite{Koopmans05,Carpene08,Krauss09,Zhang09,Koopmans10,Chimata12,Battiato10,Battiato12,Battiato14,Wienholdt2013} and are currently debated. \cite{Carva11nat,Carva11,Carva13,Eschenlohrnatmater2013,Graves13,Vodungbo12,Pfau12,SchellekensPRL13,wei_13} More recently the ultrafast build-up of magnetization in gold was measured,\cite{Melnikov11} and ferromagnetic Fe was found to undergo both an ultrafast decrease or increase of magnetization.~\cite{Rudolf12} Although this last study did not offer any direct measurement of magnetization, the reported effect was inferred from the qualitative differences observed in the experimental T-MOKE spectra at the 3p (M) absorption edge of Fe.~\cite{Rudolf12,Turgut13} For a demagnetized sample, the 3p asymmetry Fe peak is observed to decrease without noticeable changes of shape.~\cite{Rudolf12} For a sample with increased magnetization, instead, the aforementioned peak remains approximately unaltered but a shoulder grows at lower transition energies~\cite{Rudolf12,Turgut13} (shown in Fig.~\ref{fig:expdata} for convenience). While the decrease of a peak, even in the femtosecond timescale, has been already safely assigned to a net decrease of magnetisation, the growth of a shoulder is a new observation, which in Refs.~\onlinecite{Rudolf12,Turgut13} was associated to an increase of magnetization. However, this conjecture has not been fully justified yet, and a few questions remain open. Does the growth of the shoulder come from an increase of the Fe magnetization or from other effects, such as changes in the material response driven by a different element in the sample, or even by processes not involving magnetism?\cite{carvaEPL09} Answering these precise questions is an important task to understand and above all confirm the existence of an ultrafast increase of magnetization, interpret experimental data, and possibly help in designing new experiments and clarifying the very nature of the ultrafast dynamics.

The aim of the present paper is to provide a model to describe the excited states of Fe after the first picosecond of ultrafast magnetization dynamics in order to predict the magnetic response of the material in the picosecond timescale. We will demonstrate that in the picosecond timescale the system has acquired a partial thermal equilibrium that can be described using microcanonical statistics on a subspace of the whole Hilbert space. The partially-equilibrated configurations for decreased and increased magnetizations will be shown to be qualitatively different in microscopic sense, i.e. in terms of local atomic moments. Our results illustrate that after ultrafast demagnetization the system has tilted atomic magnetic moments whose lengths are equal to the equilibrium value. On the contrary, after ultrafast magnetization increase, the magnetic configuration is given by aligned atomic magnetic moments with increased lengths. This microscopic description allows us to compute dielectric tensors for both increased and decreased magnetization. The calculated spectra show the same features observed in the experimentally measured asymmetry. These results offer a strong support to the existence of an ultrafast increase of magnetization in Fe, especially if one considers that in our model increased and decreased magnetizations are treated on an equal footing from the outset. It must be emphasized that our model does not address the mechanisms driving the ultrafast magnetization dynamics, which are active in the femtosecond timescale. As a matter of fact our model shows that the details of the magnetic response are independent on these mechanisms. In a sense that will be clarified in the paper, any theory of the transient process should at the end lead to the same type of configuration. 

The article is structured as follows. After this Introduction, in Section \ref{sec:label} we will illustrate our classifications of the excited states in terms of magnetic and electronic configurations, emphasizing the boundaries of the performed approximations. In Section \ref{sec:energy} we will write explicitly the energy of our system with respect to appropriate reference states. Then, in Section \ref{sec:equilibration}, we will illustrate the mechanisms leading to the partial equilibration, in relation to the experimental timescales. In Section \ref{sec:probability}, we will clarify how to evaluate the probability of finding the system in the excited states (presented in Section \ref{sec:label}) when probing the system in the picosecond time scale. The most probable magnetic configurations in the performed approximations and under the selected constraints will be described in Section \ref{sec:mag_configuration}. The treatment of the electronic configurations for the most probable magnetic configurations will be the object of Section~\ref{sec:full_configuration}. Finally, in Section \ref{sec:response}, we will use the previous results to calculate the dielectric response, and compare it to available experimental data. Conclusions and Acknowledgements will close this manuscript.

\section{Labelling of microstates}\label{sec:label}

Any treatment of the equilibration problem through statistical mechanics requires the analysis of the Hilbert space spanned by all the excited states of the system. One should first classify all possible microstates $\Psi$ of the system, and then calculate the energy $H(\Psi)$ of every microstate. In principle one should use a quantum mechanical classification in terms of excited many-body states, but it is obvious that this treatment is infeasible for present knowledge and computational resources. Therefore, we proceed by grouping microstates according to their spin configuration, in the spirit of the adiabatic approximation suggested in Ref.~\onlinecite{antropov96}. Notice that this approximation is also one of the pillars of atomistic spin dynamics,~\cite{corina_review} but in our case does not imply a total decoupling of spin and electronic system, like e.g. in the three temperature model. We can then classify a given microstate by its magnetic configuration, i.e. the length and orientation of the local magnetic moments on the atoms of the system. We refer to a given atomic moment as ${\bf{m}}_i$ with the coefficient $i$ running from 1 to the number of atoms in the system $N_{\text{at}}$. We instead will refer to the magnetic configuration (all the atoms) as $\left\{ {\bf{m}}_i \right\}$. For simplicity we assume a material with only one atom in the unit cell; the generalisation to more atoms per cell is straightforward. To give an example, we associate to the zero temperature ferromagnetic phase with atomic moments aligned along z, the magnetic configuration $\left\{ {\bf{m}}_i=M_{\text{eq}} \hat{{\bf{z}}} \right\}$. Here and in the following, conditions within the curly brackets are assumed to be valid $\text{ } \forall i$. Moreover, $M_{\text{eq}}$ is the atomic moment length at zero temperature and $\hat{{\bf{z}}}$ is the unit vector along the $z$ direction. A generic magnetic configuration can have moments of different length and oriented in different directions. Note that, even in an itinerant ferromagnet like Fe, the d bands remain fairly localized around each nucleus and an atomic magnetic moment can be defined with very good approximation.~\cite{antropov96,rosengaard97}

\begin{figure}
\includegraphics[width=0.3\textwidth]{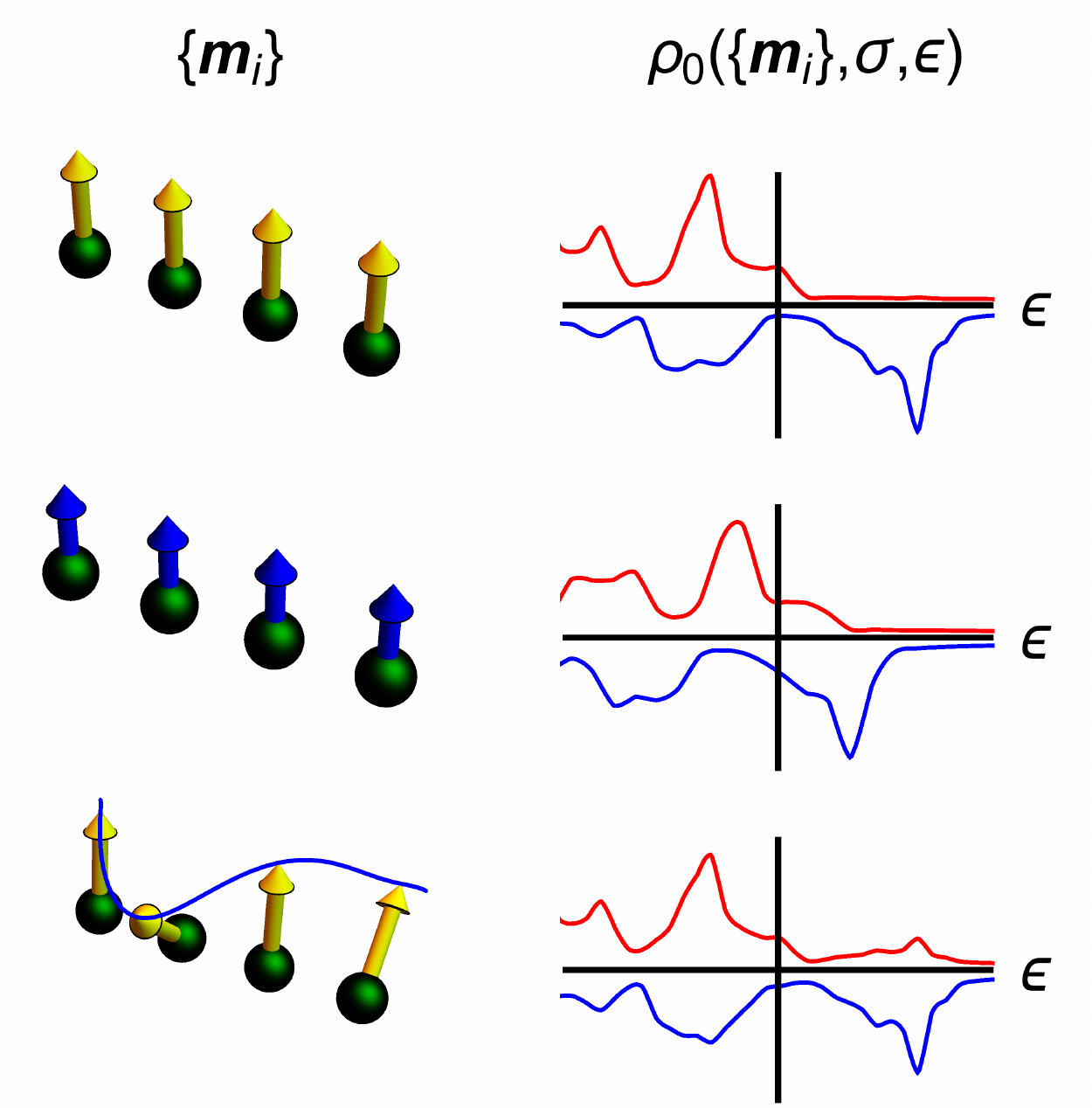}
\caption{
Some examples of rigid band structures $\rho_0(\left\{ {\bf{m}}_i \right\},\sigma,\epsilon)$ associated to given magnetic configurations $\left\{ {\bf{m}}_i \right\}$ as calculated with constrained DFT for Fe. The arrows represent the length and direction of the atomic magnetic moments, and are coloured according to their length (red, yellow and blue corresponding to moments longer, equal or shorter than $M_{eq}$, respectively). The energy dependence of  $\rho_0(\left\{ {\bf{m}}_i \right\},\sigma,\epsilon)$ is also shown, on the right side. Majority and minority spins are respectively as positive and negative values of the density of states. The zero of the energy is the Fermi energy.}\label{fig:dos_magn_conf}
\end{figure}

Providing the magnetic configuration $\left\{ {\bf{m}}_i \right\}$ alone does not univocally define the microstate $\Psi$, since the electronic degrees of freedom have not been specified yet. In fact any magnetic configuration $\left\{ {\bf{m}}_i \right\}$ identifies a group of microstates, which we refer to as a mesostate. In the following we will refer to one of these mesostates by simply specifying the magnetic configuration $\left\{ {\bf{m}}_i \right\}$. To identify a single microstate within a mesostate one needs to describe the electronic configuration. This would require to solve the electronic many-body problem for a given magnetic configuration, which makes the problem intractable. Therefore we formulate a description of all the electronic states within a mesostate with respect to the microstate with the lowest energy $\Psi_0(\left\{ {\bf{m}}_i \right\})$ for that particular mesostate. The latter can be identified for every mesostate. In particular if the magnetic configuration is $\left\{ {\bf{m}}_i=M_{\text{eq}} \hat{{\bf{z}}} \right\}$, the lowest energy microstate within the mesostate is the ground state $\Psi_{\text{GS}}$ of the system.  For a generic magnetic configuration the lowest energy state is not the ground state of the system, but can still be obtained by a constrained minimisation of the energy. In this study we have used constrained DFT, as discussed below, but other techniques may be used as well. We call the minimum energy within a mesostate $E_{\text{min}}(\left\{ {\bf{m}}_i \right\})$. We can now classify the excited states within a mesostate as superpositions of single electron promotions on the rigid band structure of the lowest energy states, as illustrated in Fig.~\ref{fig:dos_magn_conf}. In practical terms this requires first to evaluate the density of states $\rho_0(\left\{ {\bf{m}}_i \right\},\sigma,\epsilon)$ of the lowest energy microstate $\Psi_0(\left\{ {\bf{m}}_i \right\})$ in a given mesostate $\left\{ {\bf{m}}_i \right\}$, where $\sigma$ is the spin and $\epsilon$ the single electron excitation energy. Then, we can fully describe all microstates in the mesostate by specifying the electronic population of the density of states $n(\sigma,\epsilon)$, with the constraint of preserving the total number of majority and minority spin electrons. The state with the lowest energy within a mesostate is associated to the Fermi-Dirac distribution $n_F$ at zero temperature. Therefore, it is convenient to use the difference $\Delta n(\sigma,\epsilon) \equiv n(\sigma,\epsilon) - n_F(\epsilon,T=0)$ to describe the electronic repopulation of the excited states. The constraint of preserving the total number of majority and minority spin electrons now becomes simply
\begin{equation} \label{eq:chargecons}
	\int_{-\infty}^{+\infty} \!\!\!\!\!\rho_0(\left\{ {\bf{m}}_i \right\},\sigma,\epsilon)\,\Delta n(\sigma,\epsilon) d\epsilon=0 \:.
\end{equation}
This description of the excited states is fairly good in metals, provided that correlation effects are not too strong and excitation energies not too high. Correlation effects in Fe are indeed moderate~\cite{prl_igor_2009}, even when compared to other $3d$ transition metals~\cite{prb_igor_2012}. Excitation energies in the typical experimental setups we want to address are below a few hundreds of meV per atom.

\section{Hamiltonian of the system}\label{sec:energy}

\begin{figure}
 \includegraphics[width=0.49\textwidth]{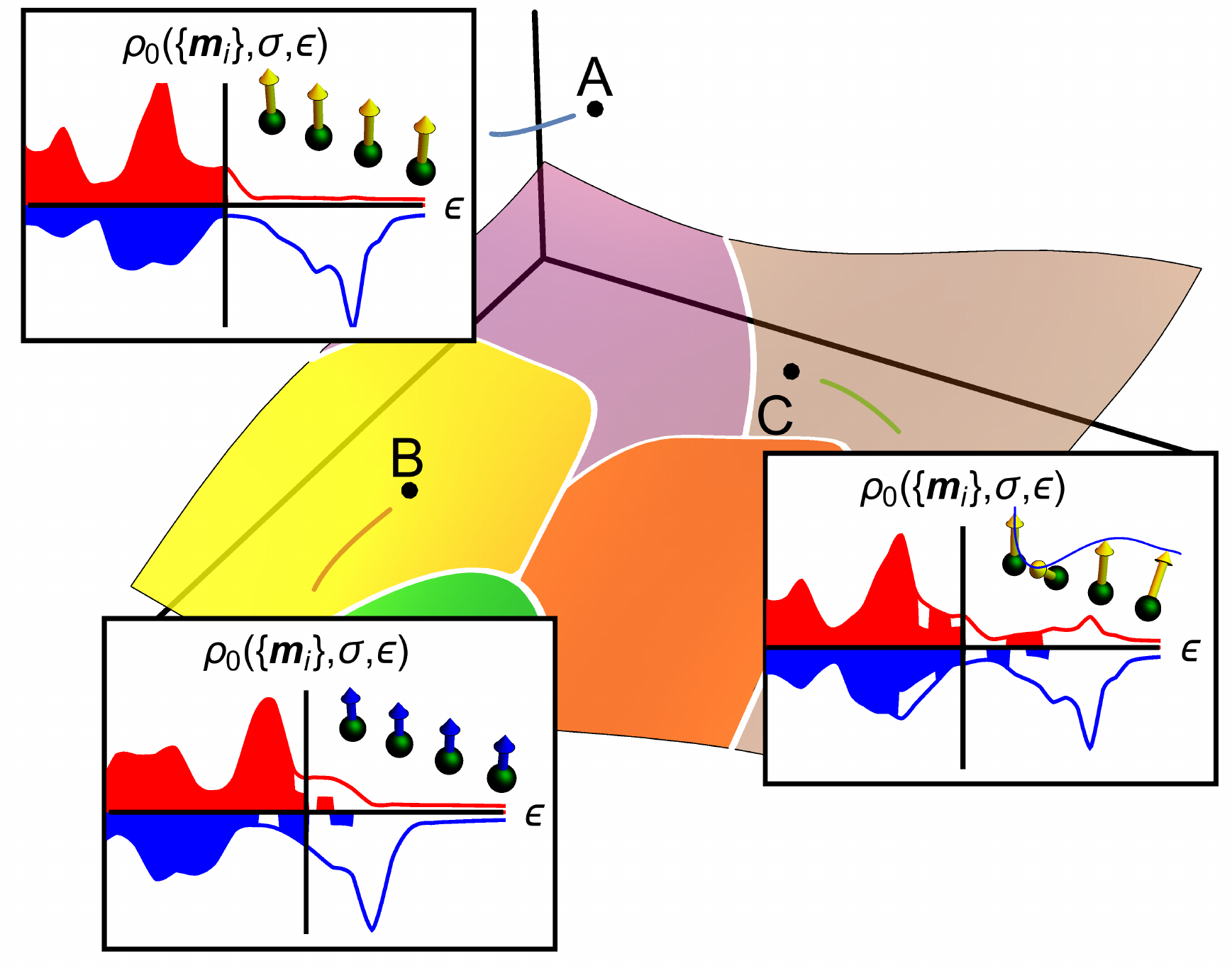}
 \caption{
 Pictorial view of the possible microstates in the Hilbert space. The surface represents the subspace satisfying the constraints of fixed energy and fixed magnetization. The coloured regions represent 
intersections of mesostates with different magnetic configurations with the constraint of fixed energy and fixed magnetization. As examples, in addition to the ground state of point A, two microstates are shown, belonging to two areas with different magnetic configurations. Point B is a microstate with atomic magnetic moments reduced in amplitude and aligned. Point C is a microstate with magnetic moments with equilibrium length but tilted directions. The insets show the density of states for both majority (red, top) and minority (blue, bottom) spins, together with a cartoon of the orientation and length of the atomic magnetic moments. The arrows are coloured according  to their length, as in Fig. \ref{fig:magnet_config}.}\label{fig:phasespace}
\end{figure}

The microstates defined in Section \ref{sec:label} identify the effective Hilbert space (see Fig.~\ref{fig:phasespace} for a pictorial view) which we will use for our statistical analysis. Before that, we must define the energy of each microstate  in terms of the magnetic configuration  $\left\{ {\bf{m}}_i \right\}$ and the electronic configuration $n(\sigma,\epsilon)$. In the approximations of Section \ref{sec:label}, the energy of a microstate  $\Psi$ can be written as:
\begin{equation} \label{eq:hamilton_generallll}
	H(\left\{ {\bf{m}}_i \right\},n(\sigma,\epsilon) ) \approx \sum_{\sigma}\int_{-\infty}^{+\infty}\!\!\! \epsilon  \,\rho_0(\left\{ {\bf{m}}_i \right\},\sigma,\epsilon)\,  n(\sigma,\epsilon)\;d\epsilon .
\end{equation}
Using the state with lowest energy within the mesostate as a reference, i.e. defining
\begin{equation}\label{eq:ham_mag}
 E_{\text{min}}(\left\{ {\bf{m}}_i \right\}) \equiv \sum_{\sigma}\int_{-\infty}^{E_F} \epsilon  \,\rho_0(\left\{ {\bf{m}}_i \right\},\sigma,\epsilon)\;d\epsilon 
\end{equation}
one can rewrite Eq.~\ref{eq:hamilton_generallll} as 
\begin{equation} \label{eq:hamilton_general}
	H(\left\{ {\bf{m}}_i \right\},n(\sigma,\epsilon) ) \approx\, E_{\text{min}}(\left\{ {\bf{m}}_i \right\}) \, + E_{\text{el}}(\left\{ {\bf{m}}_i \right\},n(\sigma,\epsilon))
\end{equation}
The second term in Eq.~\ref{eq:hamilton_general} is the contribution associated to the electronic repopulation:
\begin{multline}\label{eq:ham_ele}
 E_{\text{el}}(\left\{ {\bf{m}}_i \right\},n(\sigma,\epsilon)) \equiv  \\
   \sum_{\sigma}\int_{-\infty}^{+\infty} (\epsilon - E_F) \,\rho_0 (\left\{ {\bf{m}}_i \right\},\sigma,\epsilon)\, \Delta n(\sigma,\epsilon)\;d\epsilon \:,
\end{multline}
where $E_F$ is the Fermi energy. The equivalence between Eqs.~\ref{eq:hamilton_generallll} and \ref{eq:hamilton_general} can be promptly verified by means of Eq.~\ref{eq:chargecons}. In spite of the fact that the energy above is already derived from a few approximations, its treatment remains extremely complex due to the fact that the density of states $\rho_0(\left\{ {\bf{m}}_i \right\},\sigma,\epsilon)$ still depends in a very complex way on the full details of the magnetic configuration $\left\{ {\bf{m}}_i \right\}$. In practice, this requires the numerical calculation of $\rho_0(\left\{ {\bf{m}}_i \right\},\sigma,\epsilon)$ for almost every $\left\{ {\bf{m}}_i \right\}$. Due to the difficulties in treating directly with Eq.~\ref{eq:hamilton_general}, we approximate this expression even further, by identifying various types of mesostates.

\subsection{Moments of equal length with a large tilting}
 We focus first on the magnetic configurations where all magnetic moments are equally long ${\left\{\left| {\bf{m}}_i\right|=m \right\}}$, but may have different directions. For generic tiltings, the lowest microstate energy within the mesostate can be rewritten as the sum of the ferromagnetic reference energy for moments of arbitrary length, a Heisenberg energy which depends only on the orientation of the moments, and the anisotropy energy.~\cite{white_book} For Fe, which is the main object of this work, the anisotropy is very small if compared to the exchange,~\cite{barnstein86} and can therefore be neglected in a first approximation (see also Appendix~\ref{sec:mag_anisotropy}). We can then write:
\begin{equation} \label{eq:equal_len_large_tilt_no_anis}
\begin{split}
	E_{\text{min}}(\left\{ {\bf{m}}_i \right\}) & \approx E_{\text{min,FM}}(m) \, +  \\
	& +  \frac{1}{N_{\text{at}}} \sum_{j,k} J_{j,k} (\left\{{\bf{m}}_i\right\}) \,\left(1- \frac{{\bf{m}}_j \cdot {\bf{m}}_k}{\left|{\bf{m}}\right|^2} \right) 
\end{split}
\end{equation}
Here $E_{\text{min,FM}}(m)$ is the energy of the ferromagnetic state $\Psi_{0}({\left\{ {\bf{m}}_i=m \hat{{\bf{z}}} \right\}})$, i.e. the minimum energy obtained with the constraint of having all atomic moments aligned and with length $m$. Notice that $m$ can have an arbitrary value and be equal, smaller or bigger than $M_{\text{eq}}$. The $J_{j,k}$ are instead the intersite exchange parameters, which unfortunately for Fe depend on the full magnetic configuration $\left\{{\bf{m}}_i\right\}$.~\cite{attila_prl} This dependence is however not so drastic to reverse the sign of the exchange interaction, and therefore generating a spin wave with large tiltings will always cost a higher energy than a wave with small tiltings. This consideration is going to be sufficient for our present investigation, and it will be shown that all these mesostates contribute marginally to the statistics of the system.

\subsection{Moments of equal length with a small tilting}
For the microstates belonging to mesostates with equally long atomic moments  ${\left\{\left| {\bf{m}}_i\right|=m\right\}}$ and a small tilting between neighbours, Eq.~\ref{eq:equal_len_large_tilt_no_anis} can be simplified as:
\begin{equation} \label{eq:equal_len_tilt_no_anis}
\begin{split}
	E_{\text{min}}(\left\{ {\bf{m}}_i \right\}) & \approx E_{\text{min,FM}}(m) \, +  \\
	& +  \frac{1}{N_{\text{at}}} \sum_{j,k} J_{j,k}(m) \,\left(1- \frac{{\bf{m}}_j \cdot {\bf{m}}_k}{\left|{\bf{m}}\right|^2} \right) 
\end{split}
\end{equation}
Now the $J_{j,k}$ depend only on the value $m$, and not on the full magnetic configuration $\left\{{\bf{m}}_i\right\}$. Moreover, for small tiltings, the \textit{local} density of states of any microstate in the mesostate $\left\{ \bf{m}_i \right\}$ coincides at the leading order with the density of states obtained for all moments aligned, i.e. $\Psi_{0}(\left\{ {\bf{m}}_i=m \hat{{\bf{z}}} \text{ } \right\})$, with the only difference that the spin axis has to be rotated on every atom to align to the local moment ${\bf{m}}_i$. Therefore the electronic population of $\Psi$ can be equivalently specified on the density of states
\begin{equation}
\rho_{\text{FM}}(m,\sigma,\epsilon) \equiv \rho_0(\left\{ {\bf{m}}_i=m \hat{{\bf{z}}} \text{ } \right\},\sigma,\epsilon) \: .
\end{equation}
This greatly simplifies the second term of the Hamiltonian in Eq. \ref{eq:hamilton_general} since the density of states now depends only on the value $m$ and not on the complex details of the magnetic configuration $\left\{{\bf{m}}_i\right\}$. Note that this approximation is very good for small tiltings, corresponding to magnons of long wavelength, but fails for the opposite case. We can now rewrite Eq.~\ref{eq:ham_ele} as:
\begin{equation} \label{eq:electronic_contribution_short}
\begin{split}
	&E_{\text{el}}(\left\{ {\bf{m}}_i \right\},n(\sigma,\epsilon)) \approx \\
	&\;\;\; \sum_{\sigma}\int_{-\infty}^{+\infty} (\epsilon - E_F) \,\rho_{\text{FM}}(m,\sigma,\epsilon)\, \Delta n(\sigma,\epsilon)\;d\epsilon
\end{split}
\end{equation}
We emphasize that all terms in Eq.~\ref{eq:equal_len_tilt_no_anis} and Eq.~\ref{eq:electronic_contribution_short} can be evaluated \textit{ab initio} by means of constrained DFT,~\cite{Dederichs84} where the constraint is given by having ferromagnetically arranged atomic moments of a specified length (for details see Appendix \ref{sec:computational_details}).

\subsection{Moments of variable length with a small tilting}
Finally we need to address the generic case of magnetic configurations with  moments of variable length on neighbouring atoms. For simplicity we focus on small tilting between neighbours. For small variations of the length around an average length $\overline{m}$ we can write:
\begin{equation} \label{eq:min_different_lengths}
\begin{split}
	E_{\text{min}}(\left\{{\bf{m}}_i\right\}) \approx&  \frac{1}{N_{\text{at}}} \sum_{j,k} J_{j,k}(\overline{m}) \, \left(1- \frac{{\bf{m}}_i}{\left|{\bf{m}}_i\right|} \cdot\frac{{\bf{m}}_j}{\left|{\bf{m}}_j\right|}\right)+ \\
	& +   \frac{1}{N_{\text{at}}} \sum_{j,k} L_{j,k}(\overline{m}) \;\Big|\left|{\bf{m}}_j\right|-\left|{\bf{m}}_k\right| \Big|+\\
	&+ \overline{ E_{\text{min,FM}}}\left(\left|{\bf{m}}_i\right|\right) .
\end{split}
\end{equation}
Here the first term on the right hand side is a Heisenberg energy due to the tilting (transverse fluctuations), while the second term gives the energy increase due to magnetic moments with different lengths on different sites (longitudinal fluctuations). The last term instead represents the average of the energies for the formation of local moments of different length, i.e. 
\begin{equation}
	\overline{E_{\text{min,FM}}}\left(\left|{\bf{m}}_i\right|\right)=\sum_i E_{\text{min,FM}}\left(\left|{\bf{m}}_i\right|\right)/ N_{\text{at}} \: .
\end{equation}	
In principle, we could develop approximations to simplify the dependence of $\rho_0(\left\{{\bf{m}}_i\right\},\sigma,\epsilon)$ on the magnetic configurations but we will show that these mesostates can be neglected.

\section{Partial equilibration}\label{sec:equilibration}

\begin{figure}
\includegraphics[width=0.49\textwidth]{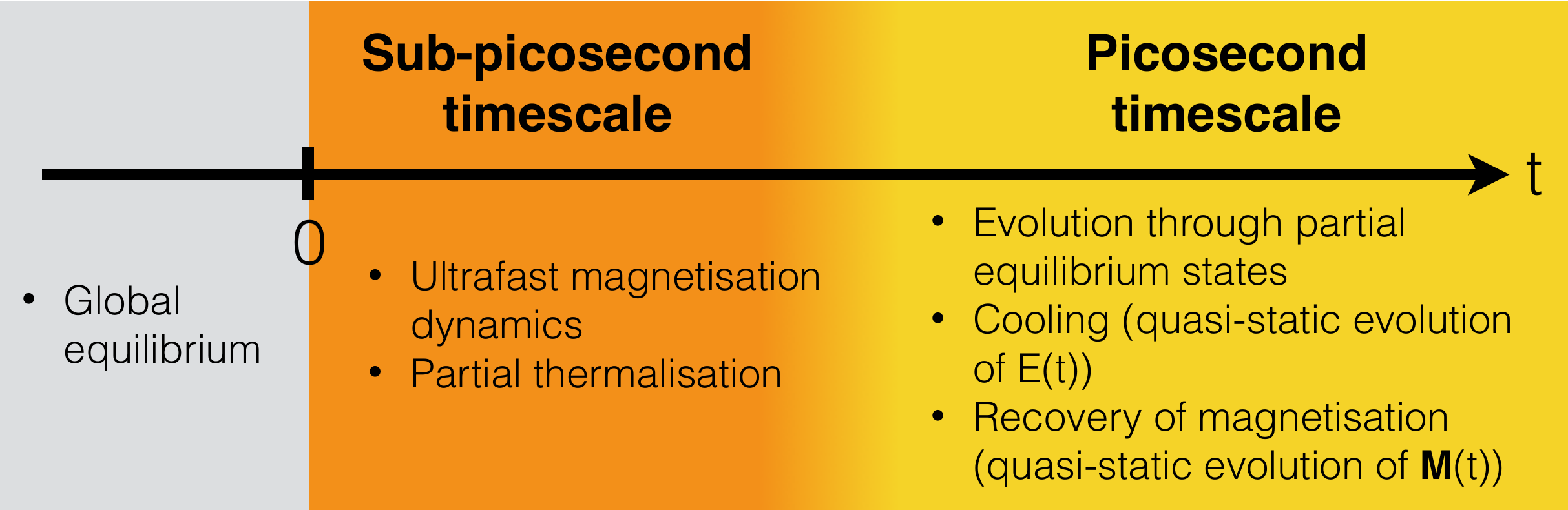}
\caption{
Schematic view of the various mechanisms active for different timescales after the laser pulse.
}\label{fig:timescales}
\end{figure}

We have so far approximated the Hamiltonian for selected microstates but did not say anything about the state of the system. In this Section we will clarify why we can treat the system as partially equilibrated and we will define the type of partial equilibration.

We split the dynamics of the system in two different timescales. For simplicity we will name them sub-picosecond and picosecond dynamics (see Fig.~\ref{fig:timescales}). We must stress that the precise estimation of the temporal length of these two types of dynamics depends on several factors, as for instance the material under study. For sub-picosecond dynamics we intend the time during which the magnetisation changes rapidly.  During this timescale the system undergoes a strong electronic excitation after the direct laser absorption. The electrons will then repopulate the density of states tending towards rebuilding a Fermi-Dirac distribution at high temperature.~\cite{Lisowski04,Lisowski05} Within the same timescale the microscopic effects which are responsible for the magnetisation dynamics will also affect the electronic configuration in some way. If the magnetisation dynamics happens before, during or after the electrons have rebuilt an internal thermal equilibrium is completely irrelevant for our discussion. What is important is that when the microscopic mechanism responsible for the ultrafast magnetisation dynamics has stopped being active, the electronic system has already attained an internal thermal equilibrium. This thermalisation happens due to the system exploring the phase space via  electron-electron (e-e) scattering. The chaotic behaviour of the electronic motion leads the system to span uniformly a part of the phase space, as we are addressing the dynamics of a closed system. For sake of simplicity, we neglect energy relaxation due to electron-phonon scattering, but its inclusion would lead to the same conclusions (see Appendices~\ref{sec:phonons} and \ref{sec:max_mesostate_prob}). It is now important to understand what is the part of the phase space that is explored via this dynamics. The main effect of the e-e scattering is the reshuffling of the energy positions of the electrons without changing the total energy of the system. The spin-orbit coupling for 3d-levels in Fe is small, which leads to a small probability of transferring spin moment to orbital moment or lattice. Therefore, in the sub-picosecond time scale, the majority of e-e scattering events preserve both the total energy $E$ and a magnetisation ${\bf{M}}=\sum_{i}{\bf{m}}_i / N_{\text{at}}$ that has been set by the microscopic mechanisms driving the ultrafast dynamics. However it should not be forgotten that these events can still lead to the transition from one mesostate $\left\{{\bf{m}}_i\right\}$ to another mesostate $\left\{{\bf{m'}}_i\right\}$, as long as the total magnetization is preserved, i.e. $\sum_{i} {\bf{m}}_i / N_{\text{at}} = \sum_{i} {\bf{m'}}_i / N_{\text{at}}$. This situation is for instance represented by the two microstates B and C in Fig.~\ref{fig:phasespace}. Notice that these transitions are fast: in itinerant ferromagnets, magnon lifetimes are usually very short (tens of femtoseconds).\cite{Zakeri2012} 

The above analysis tells us that, at the end of the sub-picosecond dynamics, the system attains a partial equilibrium, where the partial attribute is due to that this system still has a total magnetisation $\bf{M}$ which can be different (either bigger or smaller) than the global equilibrium value at that specific temperature. We can therefore describe the system by doing ensemble averages over the part of the phase space with fixed total energy $E$ and total magnetisation $\bf{M}$. It is fundamental to realise here that if spin-flip scattering events (for instance with phonons) were substantial for the thermalisation process, the system would thermalise to the full equilibrium, i.e. the system would acquire exactly the magnetisation expected at the final temperature. This is not what happens, since it is incompatible with the very existence of both the appearance of magnetisation in non magnetic materials and the increase of magnetisation in Fe. Notice that this, however, does \textit{not} exclude transfer of spin moment to the phonon system during the ultrafast magnetisation dynamics, but simply tells that this transfer cannot be of the same type as the one that leads to equilibration.

After the ultrafast change of magnetisation is finished, the thermalisation mechanism is still active but processes in the picosecond timescale become also important. Now the system undergoes different dynamics: cooling down due to heat diffusion, recovery of the magnetic moment due to the slow spin-phonon equilibration, and precession of the atomic magnetic moments in the magnetic field. The first two processes (cooling and recovery of magnetisation) can be treated as quasi static with respect to the e-e scattering. This implies that the correction to the electronic population coming from these effects can be described as a small perturbation of an associated equilibrium state with time dependent macroscopic magnetisation ${\bf{M}}(t)$ and energy $E(t)$. The precession of the atomic magnetic moments ${\bf{m}}_i(t)$, instead, leads to magnonic oscillations. Instead of focusing on ${\bf{M}}(t)$, one can repeat the discussion above directly for all ${\bf{m}}_i(t)$ and obtain the Landau-Lifshitz description of magnonic oscillations, with parameters that can be computed for the partially equilibrated state rather than for the completely equilibrated one. In this article we are only interested in the state of the system assuming ${\bf{M}}$ and $E$ at a given time $t$. We will see below that this is sufficient to describe the spectroscopy of the system at zeroth order accuracy, without the need of determining the equation of motion of ${\bf{M}}(t)$,  $E(t)$ or even ${\bf{m}}_i(t)$.

\section{Most probable mesostate}\label{sec:probability}
We are now ready to analyse the statistical mechanics of our system. As anticipated above, we are going to use the microcanonical statistics as it leads to a simpler approach. However, we stress once more that a treatment through canonical statistics is equally possible and leads to and indeed strengthens the same conclusions, as illustrated in Appendix~\ref{sec:phonons}. The fact that we focus on a closed system allows us to use the constraint of a fixed energy $E$. Moreover, as discussed in Section~\ref{sec:equilibration}, we add a further constraint on the total magnetisation ${\bf{M}}$. Notice that the magnetization as well as the energy are normalized per atom in this manuscript, unless explicitly stated. A macroscopic quantity associated to a microscopic quantity $\xi$  can therefore be evaluated as an average under the constraints of a fixed total energy $E$ and a fixed total magnetic moment ${\bf{M}}$. In more formal terms we have that the ensemble average $\left<\xi\right>$ is given by
\begin{equation}\label{eq:ens_avg} 
	\left<\xi\right> = {\textstyle \sum_{\xi}} \: \xi\, P\left(\xi|E,{\bf{M}}\right)
\end{equation}
where the sum runs over all possible values of $\xi$ and $P\left(\xi|E,{\bf{M}}\right)$ is the probability of the microscopic quantity having the value $\xi$ under the fixed constraints $E$ and ${\bf{M}}$. For an ergodic system (see Appendix~\ref{sec:ergodicity} for more details), the probability $P$ is proportional to the number of microstates $\mathcal{N} \left(\xi|E,{\bf{M}}\right)$ where the microscopic quantity, the energy and the total magnetisation have the specified values:
\begin{equation}\label{eq:prob_numstat}
	P\left(\xi|E,{\bf{M}}\right)=\frac{\mathcal{N} \left(\xi|E,{\bf{M}}\right)}{\sum_{\xi'} \mathcal{N} \left(\xi'|E,{\bf{M}}\right)}.
\end{equation}
Finding the quantity $\xi$ maximizing the probability in Eq.~\ref{eq:prob_numstat} means maximizing the term at the numerator, i.e. $\mathcal{N} \left(\xi|E,{\bf{M}}\right)$. If we consider a given magnetic configuration $\left\{{{\bf{m}}_i}\right\}$ as the microscopic quantity $\xi$, we can exploit that ${\bf{M}}$ depends on $\left\{{{\bf{m}}_i}\right\}$, and write:
\begin{equation}\label{eq:constr_mag}
	\mathcal{N} \left(\left\{{\bf{m}}_i\right\}|E,\bf{M}\right)= \mathcal{N} \left(\left\{{\bf{m}}_i\right\}|E\right) \delta_{\sum_i {\bf{m}}_i /N_{\text{at}} , {\bf{M}}} 
\end{equation}
where $\delta$ is the Kronecker delta, being 0 if the two arguments are different and 1 if they are equal. The equation above states the obvious fact that if a magnetic configuration consists of moments that do not sum up to the required total magnetic moment there are no microstates within the mesostate that can satisfy the constraints.

Unfortunately the calculation of $\mathcal{N}\left(\left\{{\bf{m}}_i\right\}|E\right)$ in Eq.~\ref{eq:constr_mag} is not as simple, since we need to count the number of microstates with energy $H=E$ within the mesostate  $\left\{{\bf{m}}_i\right\}$. From Eq.~\ref{eq:hamilton_general}, we see that this is equivalent to count the number of ways of distributing the energy $E-E_{\text{min}}\left( \left\{{\bf{m}}_i\right\} \right)$ among repopulations of the density of states  $\rho_0(\left\{{\bf{m}}_i\right\},\sigma,\epsilon)$. This calculation can be simplified by replacing the density of states with a constant averaged density of states, where the average is taken around the Fermi energy and over a range equal to a few times the energy per electron injected by the laser $\rho_0(\left\{{\bf{m}}_i\right\},\sigma,\epsilon)\approx \overline{\rho}_0(\left\{{\bf{m}}_i\right\},\sigma)$. Although this approximation is very reasonable for the energies involved in a typical experimental setup, the treatment of a density of states of more general shape is also possible, as illustrated in Ref.~\onlinecite{Roccia2010}. In this approximation, and for $N_{\text{at}}\rightarrow\infty$, we obtain (see Appendix \ref{sec:deriv_number_conf} for details) that the number of microstates within the mesostate is:
\begin{equation} \label{eq:microstate_general}
	\mathcal{N}\left(\left\{{\bf{m}}_i\right\}|E\right) \propto e^{N_{\text{at}}\sqrt{\overline{\rho}_0(\left\{{\bf{m}}_i\right\}) \, (E-E_{\text{min}}(\left\{{\bf{m}}_i\right\})) }}
\end{equation}
where
\begin{equation}
 \overline{\rho}_0(\left\{{\bf{m}}_i\right\}) \equiv \overline{\rho}_0(\left\{{\bf{m}}_i\right\},\uparrow) +\overline{\rho}_0(\left\{{\bf{m}}_i\right\},\downarrow) \:.
\end{equation}
The most probable mesostate can be found from Eq.~\ref{eq:microstate_general} by maximizing the product in the radicand. We notice that, as expected, the most probable mesostate is enormously more probable than any other state, due to the presence of $N_{\text{at}}$ (roughly the Avogadro's number) in the exponent. This is a great simplification since all the averages can be reduced to averages only over the microstates within the most probable mesostate. 

For an excitation of arbitrary intensity, particular care must be taken when maximizing the product $\overline{\rho}_0(\left\{{\bf{m}}_i\right\}) \, (E-E_{\text{min}}(\left\{{\bf{m}}_i\right\}))$. The most probable magnetic configuration will in general have some dependence on the total energy $E$, injected by the laser. However, for small excitations a particularly useful limit can be obtained, as illustrated in Appendix \ref{sec:max_mesostate_prob}. In this limit the most probable magnetic configuration is not dependent on the total energy $E$ but simply requires maximizing $E_{\text{min}}(\left\{{\bf{m}}_i\right\})$. This is already an important result, and we will come back to it in the conclusions.

\section{Magnetic configurations}\label{sec:mag_configuration}

\begin{figure}[t]
\includegraphics[width=0.49\textwidth]{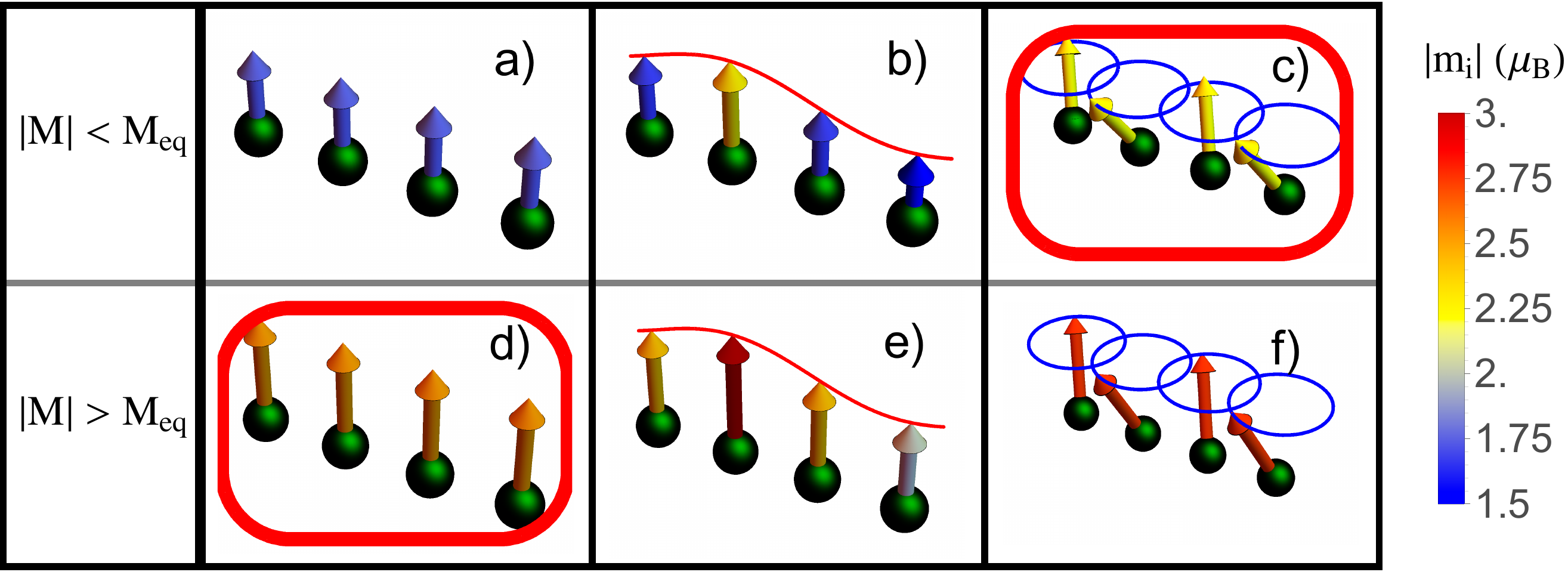}
\caption{
Magnetic configurations. The top panels show some possible magnetic configurations with a decreased average magnetic moment. In the bottom panels the same type of configurations are shown for an increased average magnetic moment. From left to right: a linear decrease or increase, amplitude spin fluctuations, transverse spin fluctuations. The arrows are coloured according to their length.}\label{fig:magnet_config}
\end{figure}

In this Section, we first identify all the possible magnetic configurations $\left\{{\bf{m}}_i\right\}$ satisfying the constraint on the total magnetic moment $ \sum_{i=1}^{N_{\text{at}}}{\bf{m}}_i / N_{\text{at}} = {\bf{M}}$. Then we will look for the magnetic configurations with the smallest $E_{\text{min}}(\left\{{\bf{m}}_i\right\})$. To cover all possible magnetic configurations it is convenient to divide them into three different groups, which are shown in Fig.~\ref{fig:magnet_config} for decreased and increased magnetization, respectively $\left|{\bf{M}}\right| < M_{\text{eq}}$  and $\left|{\bf{M}}\right| > M_{\text{eq}}$. For simplicity, we have neglected here the magnetic anisotropy and used only scalar values for the magnetization (see also Appendix~\ref{sec:mag_anisotropy}). Moreover, we have shown above that in the limit of large $N_{\text{at}}$ and constant spin-integrated averaged density of states, the energy $E$ does not change the state of the system. Therefore, the dependence on $E$ will be ignored in the following discussion.

\subsection{Decrease of magnetization}
\begin{figure}[t]
 \includegraphics[width=0.39\textwidth]{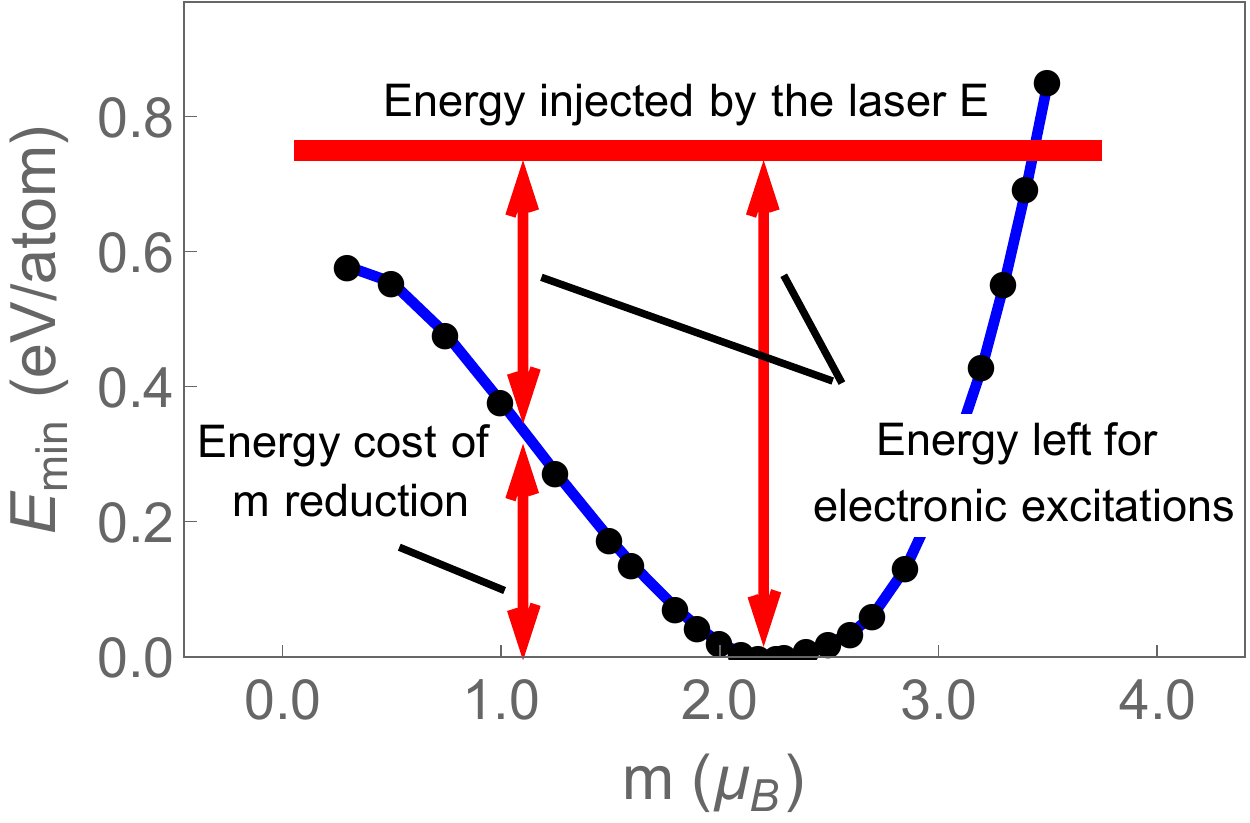}
 \caption{
 Constrained ground state energy for ferromagnetic configurations as in Fig.~\ref{fig:magnet_config}(a) and Fig.~\ref{fig:magnet_config}(d) as function of the atomic magnetic moment $m$. The zero of the energy is defined by the equilibrium atomic magnetic moment. The energy deposited by the laser is drawn schematically, and exaggerated for clarity, to emphasize what energy is left available for electronic excitations.}
\label{fig:energy_magn_mom}
\end{figure}

We first focus on a system that underwent an ultrafast decrease of magnetization, i.e. $\left|{\bf{M}}\right| < M_{\text{eq}}$. The first magnetic configuration to consider is
$\left\{ {\bf{m}}_i=M_{\text{eq}} \hat{{\bf{z}}} \right\}$, where all moments are ferromagnetically aligned and of equal (but reduced) length, as depicted in Fig.~\ref{fig:magnet_config}(a). Eq.~\ref{eq:equal_len_tilt_no_anis} shows that for no tilting $E_{\text{min}}(\left\{{\bf{m}}_i\right\})=E_{\text{min,FM}}(|{\bf{M}}|)$. This energy is considerably lower than the energy of the configuration depicted in Fig.~\ref{fig:magnet_config}(b), where all moments are ferromagnetically aligned but of different length. This can be verified by the inspection of Eq.~\ref{eq:min_different_lengths}, keeping in mind that for Fe the coefficients $L_{j,k}\left(\left\{{\bf{m}}_i\right\}\right)$ are positive and especially $E_{\text{min,FM}}(m)$ is a convex function (see e.g. Fig.~\ref{fig:energy_magn_mom}). The latter is not true if the argument $|{\bf{m}}|$ is close to zero, but this extreme case, which is anyway interesting for magnetization switching, is beyond the aim of this article. Next, we consider a magnetic configuration where the moments have a length equal to the equilibrium length $\left\{\left|{\bf{m}}_i\right|=M_{\text{eq}} \right\}$ but are tilted, as depicted in Fig.~\ref{fig:magnet_config}(c). The angles between the moments can vary but must be such to lead to the required total magnetization ${\bf{M}}$. Comparing Eq.~\ref{eq:equal_len_large_tilt_no_anis} for configurations as in Fig.~\ref{fig:magnet_config}(a) and Fig.~\ref{fig:magnet_config}(c) shows that the latter are the most favourable if the following condition is satisfied:
\begin{equation}\label{eq:spin_wave_condition}
\begin{split}
  E_{\text{min,FM}}(|{\bf{M}}|)-E_{\text{min,FM}}(M_{\text{eq}}) > \qquad  \qquad \qquad \\
 \qquad \qquad \qquad \frac{1}{N_{\text{at}}} \sum_{j,k} J_{j,k}(\left\{{\bf{m}}_i\right\}) \,\left(1- \frac{{\bf{m}}_j \cdot {\bf{m}}_k}{\left|{\bf{m}}\right|^2} \right) 
\end{split}
\end{equation}
In principle, in a sample of infinite size one can always find a spin wave of arbitrarily long wavelength satisfying the constraint on the magnetization and leading to an arbitrarily small term on the right hand side of Eq.~\ref{eq:spin_wave_condition}. If this were the only relevant mechanism for our problem, the condition in Eq.~\ref{eq:spin_wave_condition} would always be satisfied by the spin wave with the maximum wavelength allowed by the boundary conditions and compatible with the constraint. However, in practice one cannot ignore that the magnetic excitation caused by the laser pulse is initially rather localized in space, and composed of high energy magnons of small wavelength. As discussed in Section~\ref{sec:equilibration}, these magnons will quickly relax to magnons of long wavelength via magnon-magnon scattering. The relaxation time is given by the magnon lifetimes, which are smaller than a few tens of fs for short wavelength.~\cite{Zakeri2012} This means that in the picosecond timescale the state of the system is given by a superposition of spin waves of long wavelength. This in turns define a small interval of allowed lengths $\left|{\bf{m}}_i\right|$ around $M_{\text{eq}}$, if the second term on the left hand side of Eq.~\ref{eq:spin_wave_condition} is allowed to relax its argument. In principle, the precise magnetic configuration at a given time for a given material and a given laser pulse can be obtained via simulations of thermally demagnetised Fe through atomistic spin dynamics.~\cite{Chimata12}. However, this analysis is not relevant for our purposes, as we will show that the knowledge that the most probable magnetic configurations are those with $\left|{\bf{m}}_i\right| \approx M_{\text{eq}}$ and small tiltings between neighbours is sufficient for determining the dielectric response.

We finally highlight that many degenerate configurations of the shape of Fig.~\ref{fig:magnet_config}(c) can be built by symmetry. As an example, one can consider a given magnetic configuration and shift it by a lattice step. The degeneracy of these configurations implies that they are all equally probable, and must be summed over when calculating macroscopic quantities.

\subsection{Increase of magnetization}
We can now focus on the increase of magnetization, i.e. $\left|{\bf{M}}\right| > M_{\text{eq}}$, and look again for the most favourable type of configurations. For a configuration having ferromagnetically aligned moments of equal length, shown in Fig.~\ref{fig:magnet_config}(d), one obtains from  Eq.~\ref{eq:equal_len_tilt_no_anis} that  \mbox{$E_{\text{min}}(\left\{{\bf{m}}_i\right\})=E_{\text{min,FM}}(|{\bf{M}}|)$}. This energy is bigger than $E_{\text{min,FM}}(M_{\text{eq}})$ but is still the lowest value obtainable for configurations compatible with the required constraint on the magnetization. Let us look at configurations with moments that are ferromagnetically aligned but of different length, shown in Fig.~\ref{fig:magnet_config}(e). Due to the convexity of $E_{\text{min,FM}}(m)$, the extra energy required to increase some of the moments above $ |{\bf{M}}|$ is bigger than the energy gained by decreasing some other moments. In addition, there is also the energetic cost due to the coefficients $L_{j,k}\left(\left\{{\bf{m}}_i\right\}\right)$, as discussed above. Similar conclusions can be reached when focusing on a configuration where the magnetic moments have equal length but are tilted, shown in Fig.~\ref{fig:magnet_config}(f). Tilted moments must have lengths larger than $|{\bf{M}}|$ to result into an average magnetization ${\bf{M}}$, and therefore the convexity of $E_{\text{min,FM}}(m)$ leads to a higher energy. In addition, there is also an increase of energy due to the coefficients $J_{j,k}\left(\left\{{\bf{m}}_i\right\}\right)$ of Eq.~\ref{eq:equal_len_tilt_no_anis}, which makes the configuration even more costly. Therefore, the magnetic configuration minimising $E_{\text{min}}(\left\{{\bf{m}}_i\right\})$ for an increased magnetization is the one reported in Fig.~\ref{fig:magnet_config}(d), with aligned magnetic moments of increased length.

\section{Full configurations}\label{sec:full_configuration}

We have identified two qualitatively different mesostates minimizing $E_{\text{min}}$ for samples with increased and decreased magnetizations. These mesostates are so enormously more probable than the other ones that we can safely refer to them as the magnetic configurations of the system without involving the relative probabilities. Most importantly, the physical reason for the reported qualitative difference is easy to understand in our model. For reduced magnetization, the minimum energy compatible with the ferromagnetic configuration in Fig.~\ref{fig:magnet_config}(a) is given by the cost associated to the reduction of the length of the atomic moments $|{\bf{m}}_i|$, as shown in Fig.~\ref{fig:energy_magn_mom}. This is basically the intra-site exchange. On the other hand the magnetic configuration in Fig.~\ref{fig:magnet_config}(c) has an energy cost depending on both the inter-site exchange and the magnetic anisotropy energy (here ignored because of its size). The energy price for the inter-site exchange is minimized for fluctuations with a small wave-vector, and is significantly lower than the cost due to intra-site exchange. A rather different situation is observed for increased magnetization. The configuration in Fig.~\ref{fig:magnet_config}(d) has a high $E_{\text{min}}$ because of the energy needed to increase the atomic magnetic moments. However, the configuration in Fig.~\ref{fig:magnet_config}(f) has an even higher $E_{\text{min}}$ because the atomic moments, when tilted, need to be even bigger to achieve the required $|{\bf{M}}|$.

We are now left with an ensemble average over the intersection of the mesostate defined by the most probable magnetic configuration and the specific energy of the system $E$. In Sections~\ref{sec:label} and ~\ref{sec:probability}, the microstates within a mesostate were identified as all the possible repopulations $\Delta n(\sigma,\epsilon)$ of the electronic excitations in the rigid band structure $\rho_0(\left\{{\bf{m}}_i\right\},\sigma,\epsilon)$ with energy $E-E_{\text{min}}(\left\{{\bf{m}}_i\right\})$. Evaluating ensemble averages is analogous to the standard modeling of thermal averages of the response of a system. We will first identify the average population $\left<n(\sigma,\epsilon)\right>$, then note that it is enormously more probable than any other population, and finally compute the response for that population only. By considering an electronic system that has to distribute an external energy $E-E_{\text{min}}(\left\{{\bf{m}}_i\right\})$, one can obtain that the most probable population is simply the Fermi-Dirac distribution $\left<n(\sigma,\epsilon)\right>=n_F (\epsilon, T, \mu_\sigma)$ depending on three parameters, i.e. an effective temperature $T$ and two chemical potentials $\mu_\sigma$, one per each spin channel. These parameters in turn depend on the energy and the magnetic configuration through the following conditions:
\begin{align}
	\label{eq:normalization_E} & \sum_{\sigma} \int \epsilon \: n_F (\epsilon, T,\mu_\sigma) \, \rho_0(\left\{{\bf{m}}_i\right\},\sigma,\epsilon) \,d\epsilon = E\\
	\label{eq:normalization_M} & \sum_{\sigma} \int \sigma \: n_F (\epsilon, T,\mu_\sigma) \, \rho_0(\left\{{\bf{m}}_i\right\},\sigma,\epsilon) \,d\epsilon = \left|{\bf{M}}\right| \\
	\label{eq:normalization_Z} & \sum_{\sigma} \int  n_F (\epsilon, T,\mu_\sigma) \, \rho_0(\left\{{\bf{m}}_i\right\},\sigma,\epsilon) \,d\epsilon = Z,
\end{align}
where $Z$ is the electronic change per unit cell. These equations require that energy, magnetic moment and charge take the appropriate value imposed by the constraints.

\section{Dielectric response}\label{sec:response}

We are now able to compute the dielectric tensor $\boldsymbol{\varepsilon}$. The case of an increased magnetization is straightforward, as the density of states $\rho_0(\left\{{\bf{m}}_i\right\},\sigma,\epsilon)$ and the dielectric tensor $\boldsymbol{\varepsilon}$ can be obtained directly from constrained DFT calculations. The effective temperature defined by Eqs.~\ref{eq:normalization_E}-\ref{eq:normalization_Z} affects the calculations only providing a broadening, and can therefore be ignored. The case of decreased magnetisation is a bit more involved. In principle one can determine the precise magnetic configuration via atomistic spin dynamics, and then evaluate the average dielectric response of the resulting spin waves. However, a good insight into the problem can be obtained by simply using the fact that we have identified the most probable magnetic configurations as a superposition of spin waves of long wavelength. In this regime, the small tiltings between neighbouring moments have a negligible influence on the local dielectric tensor. This implies that the dielectric response of the material can be computed as an average of local responses (see Appendix~\ref{sec:mag_anisotropy} for the effect of the anisotropy), which can in turn be directly evaluated from ferromagnetic bulk Fe with a magnetization that is aligned to the local moments. As a result we have to take averages not only over many degenerate (i.e. equally probable) magnetic configurations but also spatially over the local responses. We first compute the response of a single site with a moment tilted by a given angle $\theta$ from the $\hat{{\bf{z}}}$ axis in the $zx$ direction and then rotated by an angle $\phi$ around the same $\hat{{\bf{z}}}$ axis. Calling ${\bf R}(\theta)$ and ${\bf R}(\phi)$ the two rotation matrices, we can write the dielectric response $\boldsymbol{\varepsilon}'$ as (for details refer to Appendix \ref{sec:diel_demagn})
\begin{equation}\label{eq:epsilon_prime}
	\boldsymbol{\varepsilon}'={\bf R}(\phi){\bf R}(\theta)\boldsymbol{\varepsilon}{\bf R}^{-1}(\theta){\bf R}^{-1}(\phi)
\end{equation}
In the case of Fe, and for a total magnetization directed along $\hat{{\bf{z}}}$, $\varepsilon_{xz}=\varepsilon_{zx}=\varepsilon_{yz}=\varepsilon_{zy}=0$ and $\varepsilon_{xx}=\varepsilon_{yy}\approx \varepsilon_{zz}$. 
Since the total magnetization is directed along the $\hat{{\bf{z}}}$ direction, the allowed magnetic configurations $\left\{{\bf{m}}_i\right\}$ are those where the projections of the atomic moments ${\bf{m}}_i$ in the $xy$-plane cancel out.
Therefore in any ensemble average, by symmetry, the angle $\phi$ can be integrated out. We are left with the ensemble integration of the angle $\theta$ which leads to
\begin{equation}
\left<{\boldsymbol{\varepsilon}}'\right> \approx \begin{pmatrix}
\varepsilon_{xx} &  \varepsilon_{xy} \left<\cos{\theta}\right> &   0\\ 
 -\varepsilon_{xy} \left<\cos{\theta}\right> &    \varepsilon_{xx}&  0\\ 
     0&  0& \varepsilon_{xx}
\end{pmatrix}\label{eq:lastmatrix}
\end{equation}
where $\left<\cos{\theta}\right>$ is the ensemble average of $\theta$. This can directly be linked to the ratio between the length of the average magnetic moment $\left| {\bf{M}}\right|$ and the length of the equilibrium magnetic moment length $M_{\text{eq}}$, leading simply to
\begin{equation}
	 \big<{\varepsilon_{xy}'}\big> =  \big<{\varepsilon_{xy}}\big> \frac{\left| {\bf{M}}\right|}{M_{\text{eq}}}, \;\;\;\;\; \text{if }\left| {\bf{M}}\right|<M_{\text{eq}}.
\end{equation}

We are now able to compute dielectric tensors for samples with increased and decreased magnetization. Here we focus on the off-diagonal term $\varepsilon_{xy}$, which is approximately proportional to the experimental T-MOKE asymmetry, for a system with cubic symmetry and magnetisation along $\hat{{\bf{z}}}$. These results are illustrated in Fig.~\ref{fig:exytheor}, while more detailed plots for all the components are reported in Appendix~\ref{sec:diel_demagn}. In the lower part of Fig.~\ref{fig:exytheor} we see the effect induced by a demagnetisation of the material, i.e. a simple proportional reduction of the $\varepsilon_{xy}$. Instead, the configuration with increased magnetization consists of increased atomic magnetic moments, which leads to an increased population of the spin majority band and a reduction of the population of the spin minority band. This induces a change in the density of states above the Fermi energy and an increase of the spin splitting of the core levels. As a result, the dielectric response changes only below 52 eV, as highlighted within the red boxes in Fig.~\ref{fig:exytheor}. This behavior compares qualitatively well with the experimental T-MOKE data of Refs.~\onlinecite{Rudolf12,Turgut13}, and partially reported in Fig.~\ref{fig:expdata}, showing that a shoulder grows just below the main Fe peak. This good agreement between experimental and theoretical data offers a strong theoretical support to the very existence of an ultrafast increase of magnetization.

\begin{figure}
 \includegraphics[width=0.49\textwidth]{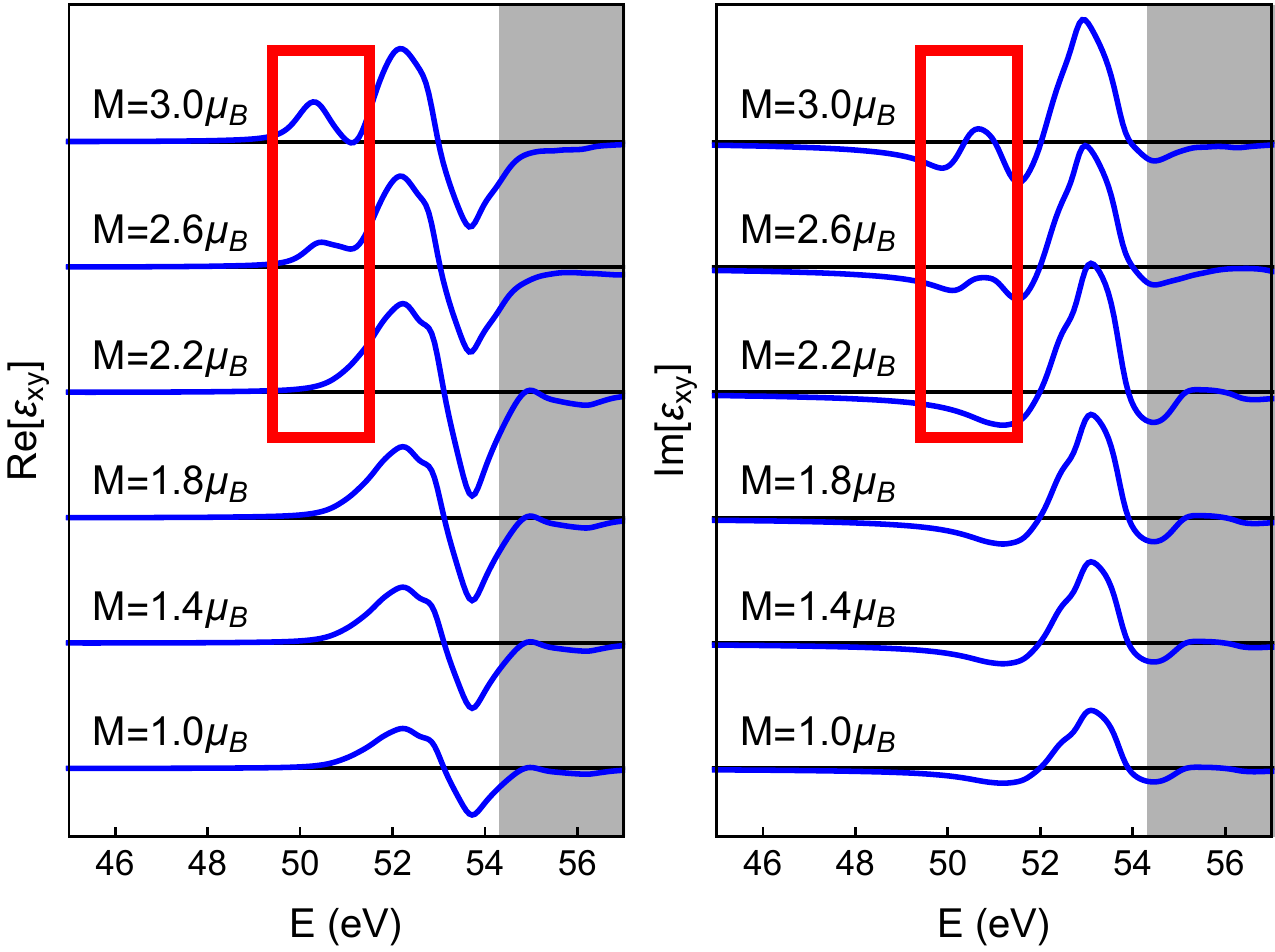}
 \caption{
Dielectric tensor element. Real and imaginary parts of the off-diagonal term of the dielectric tensor for decreased and increased magnetisations. The shaded grey area identifies those energies where the T-MOKE signals of Fe and Ni overlap, making any comparison with experimental data~\cite{Rudolf12,Turgut13} not meaningful. For increased magnetization a shoulder in the Fe peak is formed, as highlighted within the red box.}\label{fig:exytheor}
\end{figure}

\section{conclusions}

In conclusion, we provided a solid theoretical description of the microscopic states of a system right after ultrafast magnetization dynamics. Our model is based on several assumptions and approximations which reflect the high complexity of the problem under consideration. All these assumptions and approximations have been drawn on the basis of well known theoretical or experimental facts and are therefore not to be considered as a limitation of the model. In this way we are able to formulate a theory of picosecond spectroscopy. This can also be of great importance for the study of the picosecond dynamics of magnetisation, when the system is in an out-of-equilibrium state but the magnetic dynamics can be treated as a quasi-static evolution of a partially equilibrated system. 

More in particular, our model allows us to draw four major conclusions. 1) For small excitation energies the most probable magnetic configuration of the system in the picosecond timescale does not depend on energy injected by the laser. 2) For a sample of Fe, which is representative of the experimental setup in Refs.~\onlinecite{Rudolf12,Turgut13}, the state of the system right after the ultrafast demagnetisation can be described as an arrangement of tilted magnetic moments whose moduli correspond to the equilibrium magnetic moment; the state of the system right after an ultrafast increase or appearance of magnetisation can instead be described as an arrangement of aligned magnetic moments whose moduli are equal to each other but larger than the equilibrium value. 3) The dielectric response of Fe in this typical experimental setup can be calculated for the most probable magnetic configuration and a Fermi-Dirac electronic distribution over the magnetically constrained density of states. 4) The qualitative difference observed in the T-MOKE asymmetry measured experimentally for samples with decreased and increased magnetisation can be easily explained in terms of the two different states of the system. The formation of a shoulder in the T-MOKE spectrum is a feature that is also predicted by theory and can be rigorously assigned to an increase of magnetisation.

Finally, the proposed method can be applied to more general situations. Simple cases, like magnetised gold,~\cite{Melnikov11} can be treated following exactly the same arguments described in this work. More complicated cases, as for instance Gd,~\cite{Melnikov08,Carley12} require instead more care. In particular the assumption that breaks in Gd is the possibility of exploring fast enough the full space of magnetic configurations, given that d- and f-spins are expected to interact weakly with each other. We therefore expect that the partial equilibration should be done with two distinct constraints, over d-averaged and f-fixed magnetic moments. The most probable magnetic configurations are likely to be much more complicated that those proposed here for Fe. A similar approach is expected to work even in the alloys used for all-optical switching.~\cite{Stanciu07,Radu11}

\section*{Acknowledgement}
This work was sponsored by the Swedish Research Council (VR), the Royal Swedish Academy of Sciences (KVA), and the Knut and Alice Wallenberg foundation (KAW). The computations were performed on resources provided by the Swedish National Infrastructure for Computing (SNIC) at the National Supercomputer Center (NSC). The authors thank D. Rudolf, R. Chimata, O. Eriksson, J. Chico, Y. O. Kvashnin and J. Rusz for valuable discussions.

\appendix

\section{Number of configurations}\label{sec:deriv_number_conf}

For reader's convenience we report here the derivation of Eq.~\ref{eq:microstate_general}. This equation, and the following derivation, are limited to the particular case of a fermionic system with a constant density of states, and are reported only for illustrative purposes. A more general case is analyzed elsewhere,~\cite{Roccia2010} and we redirect the reader to this work for all the details. We start computing the grand canonical partition function for non interacting electrons over a constant single particle density of states. To simplify our derivation we assume this density of states to be finite only above 0, so that it has effectively the form of a step function. This assumption has no consequence on the results but makes it possible to have a much easier mathematical treatment of the interval of integration. 

We focus on a system with one spin channel, while the generalization to two spin channels is discussed below. We will first derive the expression for the discrete case of equally spaced energy levels, and we will then take the continuous limit as $N_{\text{at}}\rightarrow \infty$, where $N_{\text{at}}$ can be interpreted as the number of atoms. The discrete energy levels for one atom are defined as $E_j=j \;\delta E_1$, with $j$ being a generic quantum number going from one till infinity, and are characterized by the occupation numbers $n_j=0,1$. When going to the many atoms case, the splitting between the levels must be adjusted to be inversely proportional to the number of atoms in the system, i.e. $\delta E_{N_{\text{at}}}=\delta E_1/N_{\text{at}}$. It is easy to see that in the continuous limit the total density of states $D_{N_{\text{at}}}= 1/\delta E_{N_{\text{at}}}$ grows to infinity but the density of states per unit cell remains constant $\rho= D_{N_{\text{at}}}/N_{\text{at}}$.

The grand canonical partition function is therefore
\begin{equation}
\begin{split}
	\mathcal{Z}_N\left(\beta\right)&=\sum_{\left\{n_j\right\}}e^{-\beta\sum_{j=1}^{+\infty}j \;\delta E_{N_{\text{at}}}\,n_j + \beta \mu \sum_{j=1}^{+\infty}n_j}= \\
	&=\sum_{\left\{n_j\right\}} \prod_{j=1}^{+\infty} e^{-\beta\; j \;\delta E_{N_{\text{at}}}\,n_j + \beta \mu n_j } = \\
	&=  \prod_{j=1}^{+\infty}  \sum_{n=0,1} e^{-\beta\; j \;\delta E_{N_{\text{at}}}\,n_j+ \beta \mu n_j } =  \\
	&= \prod_{j=1}^{+\infty}  \left( 1+ e^{-\beta\; j \;\delta E_{N_{\text{at}}}+ \beta \mu}\right) \: .
\end{split}
\end{equation}
The grand potential is given by
\begin{equation}
\begin{split}
	&\Phi_G =-\frac{1}{\beta N_{\text{at}}}\sum_{j=1}^{+\infty} \ln \left( 1+ e^{-\beta\; j \;\delta E_{N_{\text{at}}}+ \beta \mu }\right) = \\
	&=- \frac{1}{\beta N_{\text{at}} \; \delta E_{N_{\text{at}}}}\sum_{j=1}^{+\infty} \ln \left( 1+ e^{-\beta\; j \;\delta E_{N_{\text{at}}}+ \beta \mu}\right) \delta E_{N_{\text{at}}}
\end{split}
\end{equation}
and in the continuous limit $N_{\text{at}}\rightarrow \infty$ one obtains that
\begin{equation}
\begin{split}
	\Phi_G &=-\frac{\rho}{\beta} \int_{0}^{+\infty}\ln \left( 1+ e^{-\beta E +\beta \mu }\right) dE = \\
	&= \frac{ \rho \, \text{Li}_2 (-e^{\beta \mu})}{\beta^2} \: ,
\end{split}
\end{equation}
where $\text{Li}_2$ is the polylogarithm (also known as Jonqui\`ere's function) of order 2.

From the grand potential it is possible to calculate the average number of fermions $<\!\!N\!\!>$  and the average energy $<\!\!E\!\!>$ at a given inverse temperature $\beta$ and chemical potential $\mu$. In the thermodynamical limit $<\!\!N\!\!>=N$ and $<\!\!E\!\!>=E$. Therefore we can write
\begin{equation}
\begin{split}
	N=\int_0^{+\infty} \frac{\rho }{1+ e^{\beta E -\beta \mu}} dE = \frac{  \rho \, \ln (1+e^{\beta \mu})}{\beta} \\
	E=\int_0^{+\infty} \frac{\rho \;E }{1+ e^{\beta E -\beta \mu}} dE = -\frac{  \rho \, \text{Li}_2 (-e^{\beta \mu})}{\beta^2} .
\end{split}
\end{equation}
We now address the case $1/\beta\ll \mu$, which means that we focus on the case for which the bottom of the band does not play a role. As stated above, this lower bound was included only to ensure a finite value of the integrals, leading to an easier mathematical treatment. It can be shown that
\begin{equation}
\begin{split}
	N&\underset{1/\beta\ll \mu}{\approx} \rho \mu \\
	E&\underset{1/\beta\ll \mu}{\approx} \frac{\rho \mu^2}{2} + \frac{\pi^2 \rho}{6 \beta^2}= E_0 + \frac{\pi^2 \rho}{6 \beta^2}\\
	\Phi_G &\underset{1/\beta\ll \mu}{\approx} - \frac{\rho \mu^2}{2} - \frac{\pi^2 \rho}{6 \beta^2}= - E_0 - \frac{\pi^2 \rho}{6 \beta^2} \,
\end{split}
\end{equation}
where we have named the minimum energy at zero temperature $E_0 \equiv \rho \mu^2/2$. We can finally focus on the entropy $S=\beta(-\Phi_G +E -\mu N )$, which can be expressed as
\begin{equation}
	S\approx\sqrt{\frac{2 \pi^2 \rho \, (E-E_0)}{3}} \: . \label{eq:entropy}
\end{equation}

Although we showed this result for a constant density of states, it holds more generally in the limit where the degenerate gas approximation holds. This is the case when the excitation energy is small compared to the bandwidth, but large compared to the level spacing around the Fermi level. In the case of a slowly varying density of states we can approximate the entropy by:
\begin{equation}
	S\approx\sqrt{\frac{2 \pi^2 \overline{\rho} \, (E-E_0)}{3}} \: ,
\end{equation}
where $\overline{\rho}$ is an average of the density of states over an energy range comparable to $E-E_0$. In the limit of a large number of atoms  the number of ways $\mathcal{N}$ to arrange the electronic excitations is proportional to $(e^S)^{N_{\text{at}}}$:
\begin{equation}
	\mathcal{N}\propto (e^S)^{N_{\text{at}}} \propto e^{N_{\text{at}}\sqrt{ \overline{\rho}   (E-E_0) }}.
\end{equation}
In the case of two spin channels, the extra energy ${E-E_0}$ that can be used for repopulating the states, can be divided over the different channels. We therefore have to integrate from the case where all energy is used by the $\downarrow$ channel though the intermediate case, until all energy is used by the $\uparrow$ channel.
The number of ways to arrange the electronic excitations becomes
\begin{equation} 
 	\mathcal{N} \propto \int_0^{E-E_0} e^{N_{\text{at}}\sqrt{\overline{\rho}_{\uparrow} \, (E') }}\;  e^{N_{\text{at}}\sqrt{\overline{\rho}_{\downarrow} \, (E-E_0-E') }} dE'
 \end{equation}
that for $N_{\text{at}}\rightarrow\infty$ simplifies to
 \begin{equation} \label{eq:num_states_magn_conf}
 	\mathcal{N} \propto  e^{N_{\text{at}}\sqrt{\left(\overline{\rho}_{\uparrow} +\overline{\rho}_{\downarrow} \right)  (E-E_0) }}.
 \end{equation}

In the main article we discuss the number of ways $\mathcal{N}$ to arrange the electronic excitations in the limit of a large number of atoms.  In the situation sketched in the main article, the minimum energy $E_0$ of the system is determined by the magnetic configuration, and is denoted as $E_{\text{min}}(\left\{{\bf{m}}_i\right\})$. The total energy of the system is denoted as $E$. This leads to the number of microstates reported in the main paper:
\begin{equation}\label{eq:num_states_rep_in_main}
\mathcal{N} \propto e^{N_{\text{at}}\sqrt{\overline{\rho}(\left\{{\bf{m}}_i\right\}) \, (E -E_{\text{min}}(\left\{{\bf{m}}_i\right\})) }}
\end{equation}
where the average density of states $\overline{\rho} (\left\{{\bf{m}}_i\right\}) $ depends on the magnetic configuration.

\section{Role of electron-phonon scattering}\label{sec:phonons}
In the main text we have assumed for simplicity that electron-phonon energy relaxation is negligible. However, its inclusion is straightforward, and does not change anything in our analysis. Again we assume that the spin-flip scattering events are not very frequent for the time scale of the equilibration. \cite{Carva13} Under this assumption, the electron-phonon interaction will make the electronic system behave according to the canonical statistics and not to the microcanonical statistics, which is used in the main paper. The constraint that the total magnetic moment remains fixed still holds. The treatment in the article can therefore be applied as it is by using the canonical statistics. The only consequence is that the averages must be evaluated on all the microstates with arbitrary total energy $E$ but weighted by the factor $\text{exp}(-\beta E)$. It means that the probability of a given magnetic configuration becomes
\begin{equation} \label{eq:canonical_statistic}
\begin{split}
   &P\left(\left\{ {\bf{m}}_i \right\}|\beta\right) \propto \\
    &\propto\int_{E_{\text{min}}(\left\{ {\bf{m}}_i \right\})}^{\infty} \!\!\!\!\!\!e^{-\beta E}e^{N_{\text{at}}\sqrt{\overline{\rho}(\left\{ {\bf{m}}_i \right\}) \, (E-E_{\text{min}}(\left\{ {\bf{m}}_i \right\})) }} \: dE.
\end{split}
\end{equation}

\section{Maximisation of the mesostate probability} \label{sec:max_mesostate_prob}

To find the most probable magnetic configuration we have to maximise $\mathcal{N}$ in Eq.~\ref{eq:num_states_rep_in_main}. This implies maximising the argument of the square root $\overline{\rho}(\left\{{\bf{m}}_i\right\}) \, (E -E_{\text{min}}(\left\{{\bf{m}}_i\right\})) $ with respect to $\left\{{\bf{m}}_i\right\}$. This maximisation is a complex problem, since it requires the \textit{ab initio} evaluation of a high number of densities of states $\overline{\rho}(\left\{{\bf{m}}_i\right\})$ and energies $E_{\text{min}}(\left\{{\bf{m}}_i\right\})$. Moreover the maximum is clearly dependent on the total energy of the system $E$.

 We show here that, however, for small excitations an analysis of the formula can lead to a substantial reduction of the complexity of the problem. We will show that the magnetic configuration that minimises $E_{\text{min}}(\left\{{\bf{m}}_i\right\}) $ is a very good approximation for the most probable magnetic configuration. 

An estimation of the error we make by taking a generic $\left\{{\bf{m}}_i\right\}$ can be written as
\begin{equation}
\begin{split}
	\mbox{err} &\propto  \frac{d \left[ \overline{\rho}\left(\left\{{\bf{m}}_i\right\}\right) \, \left(E -E_{\text{min}}(\left\{{\bf{m}}_i\right\})\right) \right]}{d \left\{{\bf{m}}_i\right\}} = \\
	& = (E -E_{\text{min}}(\left\{{\bf{m}}_i\right\})) \frac{d  \overline{\rho}(\left\{{\bf{m}}_i\right\})}{d \left\{{\bf{m}}_i\right\}}  + \\
	& \;\;\;\;\; -  \overline{\rho}(\left\{{\bf{m}}_i\right\} \frac{d E_{\text{min}}(\left\{{\bf{m}}_i\right\})) }{d \left\{{\bf{m}}_i\right\}}.
\end{split}
\end{equation}
We notice that, as the most probable magnetic configuration depends in principle on the total energy $E$, the error as well depends on the same parameter. It is clear that $E$ has to be higher than the minimum value attainable by $E_{\text{min}}(\left\{{\bf{m}}_i\right\})$. As $E$ becomes smaller and smaller, i.e.~closer and closer to the minimum value attainable by $E_{\text{min}}(\left\{{\bf{m}}_i\right\})$, the contribution to the total derivative coming from $dE_{\text{min}}(\left\{{\bf{m}}_i\right\})/d\left\{{\bf{m}}_i\right\}$ will remain unchanged, while the contribution from $d\overline{\rho}(\left\{{\bf{m}}_i\right\})/d\left\{{\bf{m}}_i\right\}$ will be more and more suppressed. In this limit the magnetic configuration $\left\{{\bf{m}}_i\right\}$ that minimises $E_{\text{min}}(\left\{{\bf{m}}_i\right\})$ is therefore an excellent approximation for the magnetic configuration that minimises the full product $\overline{\rho}(\left\{{\bf{m}}_i\right\}) \, (E -E_{\text{min}}(\left\{{\bf{m}}_i\right\})) $. It is interesting to note that in this limit, the most probable magnetic configuration does not depend on the total energy $E$.

Above we have used the microcanonical statistic. In case we want to take into account the effect of the energy relaxation with the phonons, we will have to use the canonical statistic. As described in Appendix \ref{sec:phonons}, the probability is an integral over all values of $E$, weighted by a factor $e^{\beta E}$. If $\beta$ is high only the low total energies $E$ contribute to the sum. Since all of these contributions in the integral in Eq.~\ref{eq:canonical_statistic} are approximately maximized when $E_{\text{min}}(\left\{{\bf{m}}_i\right\})$ is minimised, the same happens for their sum, i.e.~the integral.  This even strengthens the validity of the approximation to take only the magnetic configuration with the lowest energy $E_{\text{min}}(\left\{{\bf{m}}_i\right\})$ into account.

\section{Computational details} \label{sec:computational_details}
For this work we performed density functional theory (DFT) calculations, using the full-potential linearized augmented plane wave (FP-LAPW) method ELK.~\cite{elk_website} The calculations were done for body centered cubic (bcc) Fe using the experimental lattice parameter of 5.43 a.u. We addressed the ferromagnetic phase with a constrained total spin moment,~\cite{Dederichs84} where the magnetization axis was chosen to be along (100), i.e. the easy axis of Fe. The exchange-correlation functional used in DFT was the generalized-gradient approximation (GGA) by Perdew-Burke-Ernzerhof.~\cite{gga_ref} The Brillouin zone was sampled with an equally spaced grid using 30 kpoints in each direction which gave 3504 kpoints in the irreducible wedge. The muffin tin radius of the Fe spheres was set to 2.0 a.u. The basis for the valence electrons included $4s$, $4p$ and $3d$ derived states.

In order to evaluate the optical response, $3s$ and $3p$ states were added to the valence states. Spin-orbit coupling, which induces the splitting between the $3p_{1/2}$ and $3p_{3/2}$ states, was also taken into account, again with the magnetization along the easy axis. Moreover the number of empty states was converged (to a value of 40) to correctly describe the continuum of absorbing states above the Fermi level. The optical conductivity tensor was computed within linear response theory and only direct interband transitions were taken into account.~\cite{misha_ps} For the configuration with reduced magnetization, we can assume with good approximation that the local electronic structure is not modified by the formation of spin waves with a small $|\bf{q}|$. The dielectric response can therefore be computed as an average of dielectric responses of bulk systems with a rotating magnetization axis, as described in Appendix~\ref{sec:diel_demagn}.

\section{Ergodicity}\label{sec:ergodicity}
From a more theoretical point of view, the mechanism allowing isolated macroscopic quantum systems to equilibrate, the timescales involved in this process, and the details of the approach to  effective ergodicity, are under intense study.~\cite{Rigol2008,Goldstein2010,Goldstein2006,Goldstein2013,Garnerone2010,Garnerone2013} For instance, it has been argued that standard ergodicity (i.e. the identification between time and ensemble averages), given the large dimension of the Hilbert space, is not a viable mechanism to explain equilibration in isolated macroscopic quantum systems. Following a forgotten intuition by Von Neumann,~\cite{Goldstein2010} the relevant mechanism has been suggested to be the normal typicality, which is based on the most common instantaneous behaviour of the system. In particular, it has been shown that, for the vast majority of isolated (negligible interaction with the environment) quantum dynamics and for the vast majority of time instants, the microscopic state of the system is practically indistinguishable from, i.e. {\itshape{macroscopically equivalent}} to, the equilibrium state. ~\cite{Goldstein2010} The subspace of partial equilibration can now be used to evaluate the expectation value of any observable as an ensemble average. In particular we want to compute the average magnetic configuration, which, due to the macroscopical equivalence discussed above, is negligibly different from the most probable one. The most probable magnetic configuration is the one corresponding to the largest partition of the Hilbert space,~\cite{Goldstein2010} i.e associated with the highest number of microstates or equivalently the maximum of the entropy. Although the arguments above have been proved for pure states,~\cite{Goldstein2010} they can as well be applied to mixed states, as the proof is based on a density matrix formulation. Finally, we note that in the manuscript we often talk of phase space to make our arguments more intuitive, but it would be more correct to talk of a Hilbert space, due to that we are treating a quantum mechanical system.

\section{Dielectric tensor in the demagnetised state} \label{sec:diel_demagn}

\begin{figure*}[t]
 \includegraphics[width=0.9\textwidth]{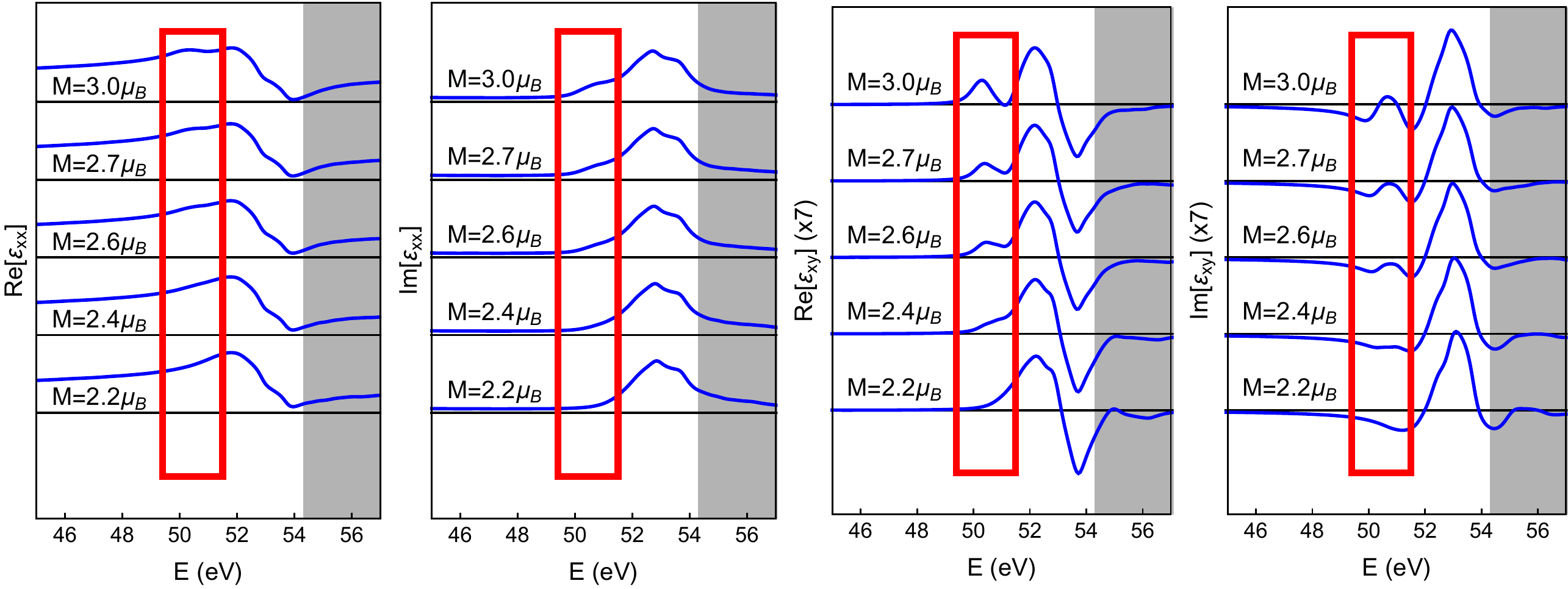}
 \includegraphics[width=0.9\textwidth]{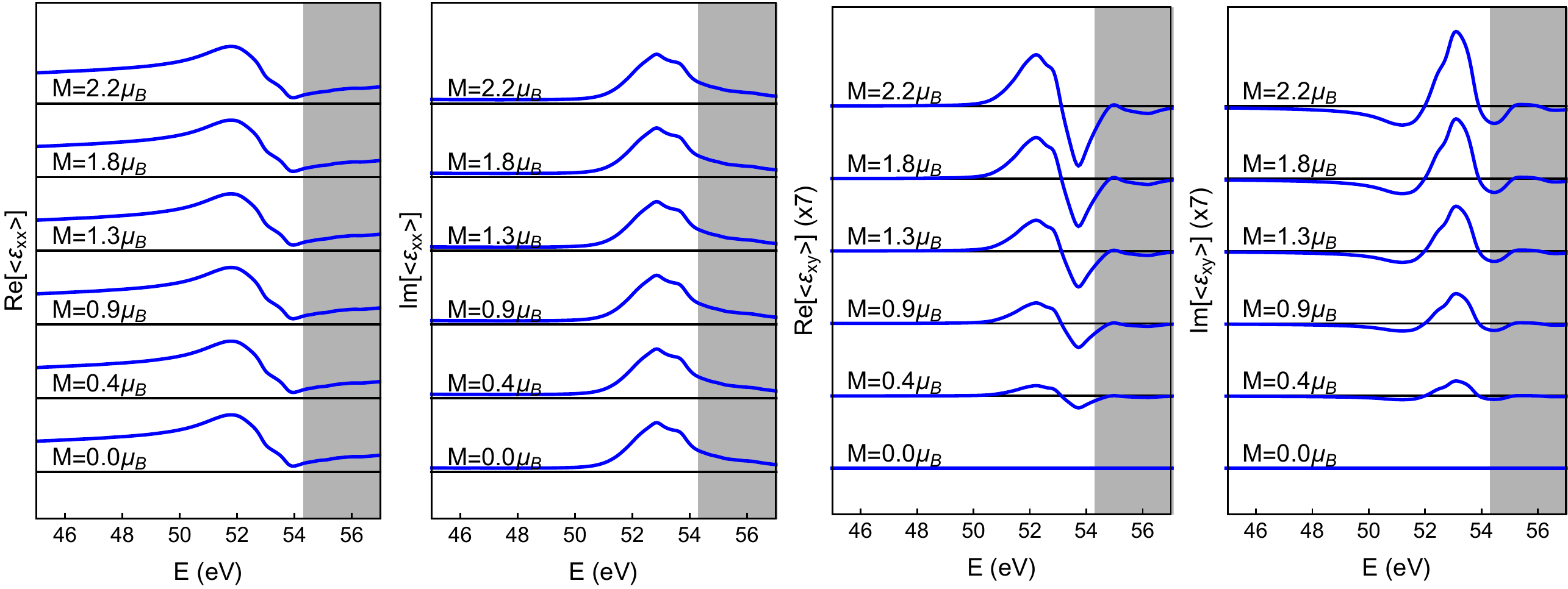}
 \caption{Diagonal and off-diagonal components of the dielectric tensor for samples of bcc Fe with increased magnetization (top) and decreased magnetization (bottom). }\label{fig:all_dielec_data}
\end{figure*}

We first tilt our magnetic moment over an angle $\theta$ from the $\hat{{\bf{z}}}$ axis. Second, we rotate our magnetic moment around the $\hat{{\bf{z}}}$ axis with an angle $\phi$ and average over all angles $0 \le \phi \le 2\pi$.
The rotation matrix over $\theta$ is given by
\begin{equation}
R(\theta)=
\begin{pmatrix}
\cos{\theta} & 0  & \sin{\theta}\\
0 & 1 & 0 \\
-\sin{\theta} & 0 & \cos{\theta}\\
\end{pmatrix}
\end{equation}
while the rotation over $\phi$ is given by
\begin{equation}
R(\phi)=
\begin{pmatrix}
\cos{\phi} & -\sin{\phi} & 0\\
\sin{\phi} & \cos{\phi} & 0\\
0 & 0& 1\\
\end{pmatrix}
\end{equation}
The tilting of the dielectric tensor by an angle $\theta$ and the following average over $\phi$ can now be rewritten as :
\begin{widetext}
\begin{eqnarray}
\left<{\boldsymbol{\varepsilon}}\right>_{\phi} &= &\frac{1}{2\pi} \int_{0}^{2\pi}{R(\phi) R(\theta) \: {\boldsymbol{\varepsilon}} \: R^{-1}(\theta) R^{-1}(\phi) d\phi} \label{epsprime}\\
&=&
\begin{pmatrix}
\frac{1}{4} (\varepsilon_{xx} + 2 \varepsilon_{yy} + 
     \varepsilon_{zz} + (\varepsilon_{xx} - \varepsilon_{zz}) \cos{2\theta})& \varepsilon_{xy} \cos{\theta} -\varepsilon_{yz}\sin{\theta}&  
  0\\ 
  -\varepsilon_{xy} \cos{\theta} + \varepsilon_{yz} \sin{\theta}&   
  \frac{1}{4} (\varepsilon_{xx} + 2 \varepsilon_{yy} + 
     \varepsilon_{zz} + (\varepsilon_{xx} - \varepsilon_{zz}) \cos{2\theta} )&  0\\ 
     0&  0&
\varepsilon_{xx} \sin^2{\theta}+\varepsilon_{zz} \cos^2{\theta}
\end{pmatrix}\nonumber
\end{eqnarray}
\end{widetext}
where the symbol $\left<\right>_{\phi}$ indicates a integration over $\phi$. 
For the case of Fe, where $\varepsilon_{xz}=\varepsilon_{zx}=\varepsilon_{yz}=\varepsilon_{zy}=0$ and $\varepsilon_{xx}=\varepsilon_{yy}\approx \varepsilon_{zz}$, this simplifies to:
\begin{equation}
\left<{\boldsymbol{\varepsilon}}\right> \approx \begin{pmatrix}
\varepsilon_{xx} &  \varepsilon_{xy} \cos{\theta} &   0\\ 
 -\varepsilon_{xy} \cos{\theta} &    \varepsilon_{xx}&  0\\ 
     0&  0& \varepsilon_{xx}
\end{pmatrix}.\label{eq:lastmatrix}
\end{equation}

\section{Complete set of results} \label{sec:complete_results}
In the main paper we have presented results only for a the off-diagonal terms of the dielectric tensor and for selected values of the magnetization. In this Appendix a more complete overview of our plots is reported. In Fig.~\ref{fig:all_dielec_data} we report all the relevant components of the dielectric tensor for the cases investigated in this work. In the top panel of Fig.~\ref{fig:all_dielec_data} data for increased magnetization are shown, where $\left|{\bf{M}}\right| > M_{eq}$. In the bottom panel, instead, data for decreased magnetization are shown, where $\left|{\bf{M}}\right| < M_{eq}$. Notice that the diagonal components do not change in case of decreased magnetization, as one can clearly see from Eq.~\ref{eq:lastmatrix}.

\section{Magnetic anisotropy}\label{sec:mag_anisotropy}
For Fe, which is the main object of this work, the anisotropy is very small if compared to the exchange,~\cite{barnstein86}, and was therefore neglected in a few equations in the main text. This was done for sake of simplicity and to avoid a text too heavy to read. In principle, Eqs.~\ref{eq:equal_len_large_tilt_no_anis},~\ref{eq:equal_len_tilt_no_anis} and \ref{eq:min_different_lengths} can be easily corrected to include the contribution of the magnetic anisotropy energy. In a first approximation this can be expressed in terms of local contributions, i.e: 
\begin{equation} \label{eq:mae} 
E_{\text{min},\text{anis}}(\left\{ {\bf{m}}_i \right\})  =  \sum_j E_{\text{anis}}({\bf{m}}_j) .
\end{equation}
This energy in Fe is of the order of the $\mu$eV, and therefore much smaller than the leading energy scales. As a result, the term in Eq.~\ref{eq:mae} can safely be ignored in our considerations on the energy minimisation leading to the most probable magnetic configuration. From a more formal point of view, including the magnetic anisotropy would add a preferred direction in the conditions of Section~\ref{sec:mag_configuration}, relating $\left|{\bf{M}}\right|$ to $M_{\text{eq}}$. In any case, the whole discussion can be repeated by using the magnetic easy axis as a reference. Finally, in the calculations of the dielectric response, the anisotropy has been neglected when evaluating $\varepsilon'$ through Eq.~\ref{eq:epsilon_prime}. Fine details of the spectrum, on the scale of the $\mu$eV may depend on this approximation, but are absolutely irrelevant for our purposes.

\bibliography{paper}


\newpage

\end{document}